\documentclass[fleqn,usenatbib]{mnras}

\usepackage{newtxtext,newtxmath,longtable}

\usepackage[T1]{fontenc}
\usepackage{pdflscape}

\DeclareRobustCommand{\VAN}[3]{#2}
\let\VANthebibliography\thebibliography
\def\thebibliography{\DeclareRobustCommand{\VAN}[3]{##3}\VANthebibliography}

\usepackage{graphicx}	
\usepackage{amsmath}	
\usepackage{float}
\usepackage{setspace}

\title[model de-idealisation]{
Spectroscopic analysis of hot, massive stars in large spectroscopic surveys with de-idealised models
}

\author[J. M. Bestenlehner et al.]{
J. M. Bestenlehner$^{1,2}$,\thanks{E-mail: j.m.bestenlehner@sheffield.ac.uk}
T. En{\ss}lin$^{3}$,
M. Bergemann$^{2}$,
P. A. Crowther$^{1}$,
M. Greiner$^{3}$,
and M. Selig$^{4}$
\\
$^{1}$Department of Physics \& Astronomy, Hounsfield Road, University of Sheffield, S3 7RH, UK\\
$^{2}$Max Planck Institute for Astronomy, K\"onigstuhl 17, 69117 Heidelberg , Germany\\
$^{3}$Max Planck Institute for Astrophysics, Karl-Schwarzschildstra{\ss}e 1, 85748 Garching, Germany\\
$^{4}$DBFZ Deutsches Biomasseforschungszentrum gemeinn\"utzige GmbH, Torgauer Stra{\ss}e 116, 04347 Leipzig, Germany
}

\date{Accepted XXX. Received YYY; in original form ZZZ}

\pubyear{2023}

\begin{document}
\label{firstpage}
\pagerange{\pageref{firstpage}--\pageref{lastpage}}
\maketitle

\begin{abstract}
Upcoming large-scale spectroscopic surveys with e.g. WEAVE and 4MOST will provide thousands of spectra of massive stars, which need to be analysed in an efficient and homogeneous way. Usually, studies of massive stars are limited to samples of a few hundred objects which pushes current spectroscopic analysis tools to their limits because visual inspection is necessary to verify the spectroscopic fit. Often uncertainties are only estimated rather than derived and prior information cannot be incorporated without a Bayesian approach. In addition, uncertainties of stellar atmospheres and radiative transfer codes are not considered as a result of simplified, inaccurate or incomplete/missing physics or, in short, idealised physical models.


Here, we address the question of ``How to compare an idealised model of complex objects to real data?'' with an empirical Bayesian approach and maximum a {\it posterior} approximations. We focus on application to large scale optical spectroscopic studies of complex astrophysical objects like stars. More specifically, we test and verify our methodology on samples of OB stars in 30 Doradus region of the Large Magellanic Clouds using a grid of FASTWIND model atmospheres.

Our spectroscopic model de-idealisation analysis pipeline takes advantage of the statistics that large samples provide by determining the model error to account for the idealised stellar atmosphere models, which are included into the error budget. The pipeline performs well over a wide parameter space and derives robust stellar parameters with representative uncertainties.

\end{abstract}




\section{Introduction}

With the advent of large spectroscopic surveys using instruments such as WEAVE \citep{2022arXiv221203981J} and 4MOST \citep{deJong2019} tens of thousands spectra of massive stars ($\gtrsim 10 M_{\odot}$) will be obtained, which will need to be analysed in a homogeneous and efficient way \citep[e.g.][]{cioni2019, 2019Msngr.175...35B, 2019Msngr.175...30C}. Current pipelines of large spectroscopic surveys are largely designed for FGK stars, which are either data driven \citep[e.g.][]{2015ApJ...808...16N, 2020A&A...644A.168G} or model driven \citep[e.g.][]{2015AAS...22542207A, 2019ApJ...879...69T}. Traditionally, and still widely performed today, massive stars have been analysed by ``eye'', which limits the sample size to $< 100$ massive stars. In addition, stellar parameters as well as uncertainties are estimated rather than determined. Larger sample of a couple of hundreds of stars are usually analysed with a $\chi^2$-minimisation algorithm, where the final fit often needs to be visually verified depending on the goodness-of-fit. 

Multi-dimensional probability distribution functions are obtained depending on the number of free parameters and uncertainties are then defined on confidence intervals rather than Gaussian standard deviations. Those uncertainties can be highly asymmetric and very large in the case of degenerated parameters. In the massive star community there are 2 main flavours of $\chi^2$-minimisation algorithms, grid based \citep[e.g.][]{simon-diaz2011, castro2012, bestenlehner2014} and Genetic Algorithms on the basis of natural selection \citep[e.g.][]{mokiem2007, brands2022}. All those algorithms use a pre-defined selection of spectral line regions for their analysis.


Theoretical models of complex physical systems are necessarily idealisations. The implied simplifications allow us to focus the view on the essential physics, to keep the model computationally feasible, and to investigate systems for which not all of their components are perfectly known. In contrast to solar-like stars, massive stars have strong stellar winds, which influence the structure of the stellar atmospheres (line blanketing) and a pseudo photosphere at optical depth 2/3, which is located in the stellar winds. The inclusion of line-driven winds into stellar atmosphere models requires the assumption of spherical geometry in the co-moving frame of the star (expanding atmosphere). In addition to effective temperature and surface gravity, mass-loss rate, velocity law, terminal velocity, wind inhomogeneity and line blanketing need to be included into the stellar atmosphere code. The stellar atmosphere of massive stars significantly depart from local thermal equilibrium (LTE) and therefore must be computed in fully non-LTE, which is computational expensive \citep[e.g][]{santolaya1997, hillier1998, graefener2002}. This limits state-of-the-art stellar atmosphere codes for hot, massive stars to 1D. 

When faced to real data of the actual system these models often perform insufficiently when judged on a goodness-of-fit basis. The difference between real and modelled data can easily exceed the error budget of the measurement. The reason is that the idealized models do not capture all aspects of the real systems, but they are still present on the data. To discriminate these missing aspects from measurement errors or noise, we use the term \textit{model errors} to describe these imperfections of our theoretical description \citep{2018arXiv181208194O}.

Often one tries to determine the model parameters from the data via a likelihood based methodology such as $\chi^{2}$-minimisation, maximum likelihood or Bayesian parameter estimation that uses the measurement uncertainties as a metric in data space. The resulting parameter estimates can be strongly distorted by the desire of the method to minimize all apparent differences between predicted and real data, indifferently if these are due to measurement or model errors.

Thus the model errors should be included into the error budget of the parameter estimation. This would require that we have a model for the not yet captured aspects of the system or at least for the model errors these produce. To do this thoroughly we would need to undertake a case by case analysis of the missing physics. 

However, this would be quite impractical in cases of the spectroscopic analysis of complex astrophysical objects. Instead, we want to construct a plausible, but by no means perfect, effective description of the model errors. In the construction of the de-idealisation model, we will follow a pragmatic route, but try to indicate the assumptions, approximations and simplifications made.

In Section~\ref{s:method} we introduce the methodology, which is used in our spectroscopic analysis pipeline (Section~\ref{s:sap}). Using grids of stellar atmospheres (Section~\ref{s:sag}) the pipeline is applied to observational data (Section~\ref{s:od}). The results are discussed in Section~\ref{s:re_di}. We close with a brief conclusion and outlook (Section~\ref{s:con_out}).  

\section{Method}\label{s:method}
\subsection{Data model}

We assume we have a set of objects (e.g. stars, galaxies, ...: labelled by $i\in\{1,\,2,\,\ldots\mathtt{n}\}$) with observable signals $s^{(i)}=(s_{x}^{(i)})_{x}$ over some coordinate $x$, e.g. the emitted spectral energy distribution $s^{(i)}=(s_{\lambda}^{(i)})_{\lambda}$ as a function of the wavelength $\lambda$. These signals are measured with a linearly responding instrument (response matrix $R$) with additive Gaussian noise ($n$) according to the measurement equation

\begin{equation}
d^{(i)}=R^{(i)}s^{(i)}+n^{(i)}.\label{eq:measurement model}
\end{equation}
 The individual elements of the data vector ($d$) for the $i$-th object
are then given by
\begin{equation}
d_{j}^{(i)}=\int dx\,R_{j\,x}^{(i)}s_{x}^{(i)}+n^{(i)},
\end{equation}
where in our spectroscopic cases $R_{j\,x}^{(i)}$ is the $j$-th bandpass of our $i$-th observation as a function of wavelength $x=\lambda$. Spectroscopic, colour filter, and bolometric measurements can thereby be treated with the same formalism and even combined into a single data vector and response matrix. In addition, we do not require that all objects are observed in the same way by keeping the response matrix dependent on the object index $i$. In this way, the formalism permits to combine heterogeneous observations. 

For the measurement noise $n^{(i)}$ of the $i$-th observation we use the error-spectrum from the data reduction which is assumed for simplicity to be Gaussian with zero mean and signal independent, 
\begin{eqnarray}\label{eq:gnm}
\mathcal{P}(n^{(i)}|s^{(i)}) & = & \mathcal{G}(n^{(i)},N^{(i)})\\
 & = & \frac{1}{\sqrt{|2\pi N^{(i)}|}}\exp\left[-\frac{1}{2}n^{(i)\dagger}\left(N^{(i)}\right)^{-1}n^{(i)}\right]\nonumber 
\end{eqnarray}
with assumed noise covariance $N^{(i)}=\langle n^{(i)}n^{(i)\dagger}\rangle_{(n^{(i)}|s^{(i)})}$. The dagger denotes the transposed and complex conjugated vector. The noise of the different observations is assumed to be independent as well, $\mathcal{P}(n|s)=\mathcal{G}(n,N)=\prod_{i}\mathcal{G}(n^{(i)},N^{(i)})$ with $n=(n^{(i)})_{i}$ and $s=(s^{(i)})_{i}$ being the combined noise and signal vectors.

\subsection{Model errors}\label{s:me}

Now we assume that some idealised model for our objects exists that predicts a specific theoretical signal $t^{[p]}$ given a set of unknown model parameters $p$. These parameters should be physical quantities like surface gravity, radius, effective temperature, stellar wind properties or chemical composition of the object, so that well defined values $p^{(i)}$ exist for each object\footnote{Counter example would be purely phenomenological parameters, describing aspects of the data that contain observation dependent properties, or such that only make sense within a specific object description methodology. Although the here proposed approach might be applicable to such phenomenological descriptions as well, we currently demand the physical existence of the used parameters in order to be on epistemologically firm grounds.}. Those idealised models can be generated with stellar atmosphere and radiative transfer codes. Knowing these parameters for each object is the primary goal of the inference. In principle, the relation between parameters and signal could be stochastic, but for simplicity we concentrate on deterministic models. The de-idealisation model we develop should serve as an effective description for the remaining stochasticity.

The idealized model captures hopefully the dominant properties of the system but certainly not all aspects. Therefore the real signal $s^{(i)}$ of an object will deviate by an unknown stochastic component $u^{(i)}$, the model error, so that 
\begin{equation}\label{eq:mdi}
s^{(i)}=t^{[p_{i}]}+u^{(i)}.
\end{equation}
The aim of model de-idealization is to find an appropriate stochastic model $\mathcal{P}(u|p)=\prod_{i}\mathcal{P}(u^{(i)}|p^{(i)})$ for the model errors. With such, the likelihood becomes

\begin{equation}
\mathcal{P}(d|p)=\int\mathcal{\!D}u\,\mathcal{P}(d|s=t^{[p]}+u)\,\mathcal{P}(u|p),\label{eq:u-marginalization}
\end{equation}
where $\int\!\mathcal{D}u$ denotes a phase space integral for the model errors. 

In our case, we want to restrict ourselves to using the simplest possible representation of the model uncertainties. This means that we take only the first and second moments of $u^{(i)}$ into account, $v^{(i)}=\langle u^{(i)}\rangle_{(u|p)}$ and $U^{(i)}=\langle(u-v)^{(i)}(u-v)^{(i)\dagger}\rangle_{(u|p)}$, and assume the fluctuations of different objects to be independent, $\langle u^{(i)}u^{(j)\dagger}\rangle_{(u|p)}=v^{(i)}v^{(j)\dagger}+\delta_{ij}U^{(i)}$. The probability distribution that represents mean and variance without further information on higher order correlations naturally is a Gaussian with this mean and variance. Among all possible probability distributions with given mean and variance it has a maximal entropy \citep[e.g.][]{2003prth.book.....J,2008arXiv0808.0012C}. By adopting a Gaussian for the model errors,
\begin{equation}
\mathcal{P}(u|p)=\mathcal{G}(u-v,\,U),
\end{equation}
the least amount of spurious information is entered into the inference system in case only $v$ and $U=\langle(u-v)\,(u-v)^{\dagger}\rangle_{(u|p)}$ are considered. 

This does not mean that the model error statistics is a Gaussian in reality. It just means that higher order correlations are ignored for the time being. Taking such higher order correlations into account would most certainly improve the method, but is left for future work.

The Gaussianity of measurement noise and modelling error description permits us to integrate Equation~\eqref{eq:u-marginalization} analytically leading to 

\begin{eqnarray}
\mathcal{P}(d|p,\,v,\,M) & = & \mathcal{G}(d-R(v+t^{[p]}),\,M),\label{eq:effektive measurement model}
\end{eqnarray}
with $M=N+R\,U\,R^{\dagger}$ being the combined error covariance.

\subsection{Implicit hyperprior}\label{s:ih}

The de-idealisation model requires that the auxiliary parameters $v$ and $U$ are determined as well, or better marginalised over. This requires that we specify our prior knowledge on these parameters, $\mathcal{P}(v,U|p)$, which is a very problem specific task. Using such a hyperprior we could then derive an auxiliary parameter marginalised estimator for our desired model parameters $p$. A good, but numerically very expensive approach would be to sample over the joint space of model and auxiliary parameters, $p$, $v$, and $U$, for example using the Gibbs sampling method \citep[e.g.][]{2004PhRvD..70h3511W,2010MNRAS.406...60J}.

In order to have a generic, affordable and pragmatic method we introduce a number of approximations and simplifications. The first is that we replace the auxiliary parameter marginalisation by an estimation using the following approximation

\begin{eqnarray}
\mathcal{P}(d,p) & = & \int\!\mathcal{D}v\int\mathcal{D}U\,\mathcal{P}(d|p,\,v,\,U)\:\mathcal{P}(v,U|p)\,\mathcal{P}(p)\nonumber \\
 & \approx & \mathcal{P}(d|p,\,v^{\star},\,U^{\star})\,\mathcal{P}(p)
\end{eqnarray}
with $v^{\star}$ and $U^{\star}$ being suitable estimates of the auxiliary parameters and $\mathcal{P}(p)$ the parameter prior. Instead of constructing these point estimators using the so far unspecified and problem specific priors we propose to pragmatically specify them with an educated ad-hoc construction. 

The idea is to assume for a moment that a correct model parameter classification $p^{(i)}$ for any object exists, which has later on to be estimated self consistently with all the other estimates via iteration. The difference of the signals reconstructed from the data $m^{(i)}=\langle s^{(i)}\rangle_{(s^{(i)}|d^{(i)},\,p^{(i)})}$ and the one predicted from the model $t^{[p^{(i)}]}$ plus the current guesses for $v^{\star}$,
\begin{equation}
\text{\ensuremath{\delta}}^{(i)}=m^{(i)}-t^{[p^{(i)}]}-v^{(i)\star}
\end{equation}
can be analysed to provide information on $v$ and $U$. The signal reconstruction can be done via a Wiener filter, since this is optimal in case of a linear measurement and Gaussian noise model Equation~\eqref{eq:effektive measurement model}. For the signal difference this is 
\begin{eqnarray}
\!\!\!\!\!\!\!\!\delta^{(i)} & \!\!= & \!\!D^{(i)}R^{(i)\dagger}\left(N^{(i)}\right)^{-1}\!\left[d^{(i)}-R^{(i)}\left(t^{[p^{(i)}]}+v^{(i)\star}\right)\right]\!\!,\!\!\!\!\!\!\!\!\!\!\nonumber \\
\!\!\!\!\!\!\!\!D^{(i)}\!\! & =\!\! & \left[\left(U^{(i)\star}\right)^{-1}+R^{(i)\dagger}\left(N^{(i)}\right)^{-1}R^{(i)}\right]^{-1},\label{eq:WF}
\end{eqnarray}
where $D^{(i)}$ is the Wiener variance or uncertainty of the reconstruction. For the current case we have used some guesses for $v^{\star}$ and $U^{\star}$ that need to be updated accordingly to the information contained in the statistics of the signal differences $\delta^{(i)}$. To do so, we introduce a suitable proximity measure in parameter space, $\omega_{ii'}=\mathrm{prox}(p^{(i)},p^{(i')})$, that indicates for an object $i$ how much another objects $i'$ can be used to learn about the model error statistics of $i$. A naive choice would be $\omega_{ii'}=1$ always, assuming that the model error statistics is everywhere the same in the model parameter space. A more sophisticated method would partition the parameter space into characteristic regions (e.g. corresponding to the different known star and galaxy classes in our spectroscopic example) and to $ $set $\omega_{ii'}=1$ or $\omega_{ii'}=0$ in case $i$ and $i'$ belong to the same or different classes. Even more sophisticated proximity weighting schemes can be imagined with $\omega_{ii'}=1/(1+\mathrm{dist}(p^{(i)},p^{(i')}))$ using some distance measure in parameter space. However, in our case we group objects together with respect to their main line diagnostics and analyse them in the same batch.

Given such a scheme, the update operation for $v^{\star}$ and $U^{\star}$ are
\begin{eqnarray}
v^{(i)\star} & \rightarrow & v^{(i)\star}+\sum_{i'}\frac{\omega_{ii'}}{\Omega_{i}}\delta^{(i')},\nonumber \\
U^{(i)\star} & = & \sum_{i'}\frac{\omega_{ii'}}{\Omega_{i}}\left(\text{\ensuremath{\delta}}^{(i')}\text{\ensuremath{\delta}}^{(i')\dagger}+D^{(i')}\right)\mbox{ with}\label{eq:generalizedcritical filter}\\
\Omega_{i} & = & \sum_{i'}\omega_{ii'}.
\end{eqnarray}
The $v^{\star}$ update operation is a simple absorption of any systematic difference into the mean component of the model error $v^{\star}$. For an initial step it is better to set $v^{(i)\star}_{\mathrm{iteration\#0}}=0$ as it would absorb any differences between data and model, even though the model might be not representative of the objects. However, we find that the the best choice for $v$ is to represent non-stellar features, which are not part of the model, like nebular lines, interstellar bands or telluric lines.

The $U^{\star}$ update incorporates the variance in the signal difference reconstructions as well as their Wiener variances. The latter express the level of missing fluctuations of the Wiener filter reconstruction. The correction with the Wiener variance is done in analogy to the critical filter methodology developed in \citep{2011PhRvD..83j5014E,2010PhRvE..82e1112E}, where a similar variance estimation was derived under a non-informative prior on the power spectrum of a statistical homogeneous random process. For this study we intuitively extended this to non-diagonal covariance structures. This means we adopt implicitly a non-informative hyperprior for the model error statistics as summarised by $v$ and $U$. 

The logic behind this implicit prior is as follows. Assuming we would have managed to specify an appropriate non-informative hyperprior for $v$ and $U$. From this we would derive some recipe for our point estimates $v^{\star}$ and $U^{\star}$ using some approximations. The resulting recipe should not incorporate hidden spurious assumptions. The $v^{\star}$ and $U^{\star}$ estimates therefore can only be build from elements like $\text{\ensuremath{\delta}}^{(i')}$, $\text{\ensuremath{\delta}}^{(i')}\text{\ensuremath{\delta}}^{(i')\dagger}$, and $D^{(i')}$, the latter being a summary of the former elements. Requiring the estimators to be unbiased and to enclose the mentioned critical filter as a limiting case, which is an unbiased power spectrum estimator, then fixes the numerical coefficients in front of $\text{\ensuremath{\delta}}^{(i')}$, $\text{\ensuremath{\delta}}^{(i')}\text{\ensuremath{\delta}}^{(i')\dagger}$, and $D^{(i')}$ in Equation~\eqref{eq:generalizedcritical filter} to unity.

We admit that there are some frequentist elements in this derivation, since it postulates an estimator and argues for its appropriateness using bias arguments. We hope that it will be replaced by a more stringent calculation once a suitable non-informative hyperprior has been specified. For the time being, we use it in iteration with model parameter estimation.

\subsection{Method summary}\label{s:meth_sum}

The combined de-idealized model parameter estimation method comprises the following steps:
\begin{enumerate}
\item Specify the weighting scheme $\omega_{ii'}=\mathrm{prox}(p^{(i)},p^{(i')})$ that determines how similar the model parameters of two objects have to be so that their model error statistics can be assumed to be similar.
\item Adopt some initial guess or default values for the model parameters $p^{(i)}$ and mode error parameters $v^{(i)\star}$ and $U^{(i)\star}$. A naive choice could be $p^{(i)}=p$ for some plausible central $p$ within the physically permitted parameter space, $v^{(i)\star}=0$, and $U^{(i)\star}=\sum_{i'}\left(R^{(i')\dagger}d^{(i)}-t^{[p]}\right)\left(R^{(i')\dagger}d^{(i)}-t^{[p]}\right)^{\dagger}$. 
\item Calculate $\delta^{(i)}$ and $D^{(i)}$ according to Equation~\eqref{eq:WF} for all objects.
\item Update $v^{(i)\star}$ and $U^{(i)\star}$ according to Equation~\eqref{eq:generalizedcritical filter} for all objects.
\item Update the model parameters $p^{(i)}$ of all objects using the combined likelihoods Equation~\eqref{eq:effektive measurement model} that incorporates measurement and model errors. 
\item Repeat steps 3-5 until convergence. 
\end{enumerate}
The resulting estimate will be similar to a joint maximum a posteriori estimation of the model and model error parameters which are known to perform worse than a properly marginalised with respect to the model error parameter posterior mean estimator. However, given the large number of degrees of freedom in the signal space (e.g. a highly resolved emission spectrum of a star or galaxy), such optimal estimators can be extremely expensive computationally. The proposed method might therefore be favourable in many circumstances, despite its approximative and partly ad hoc nature.

In order to be explicit about the assumptions and approximations adopted we provide an overview below. This list should help to judge the range of applicability and to find possible improvements of the proposed method. In particular we assume
\begin{enumerate}
\item that the measurement noise is independent for the different objects and independent of their signals, it has Gaussian statistics with known covariance.
\item a linear and known measurement response.
\item that the model error knowledge can be approximated by a multivariate Gaussian in signal space.
\item that regions in parameter space exist and are known which have similar model error statistics as parametrized by a mean and variance. 
\item no prior knowledge on the values of this model error mean and variances.
\item that an iterated point estimate of all involved parameters leads to a reasonable approximative solution of the joint inference problem. 
\end{enumerate}

\section{Spectroscopic analysis pipeline}\label{s:sap}

The pipeline has been developed using {\it python3} with commonly used libraries such as {\it numpy}, {\it scipy}, {\it pandas}, {\it multiprocessing} and {\it ctypes} plus {\it astropy.io} to read fits files. Using commonly and maintained libraries will ensure that the pipeline is easy to maintain and should be usable over a long period of time. The following section outlines the required pre-processing steps (\S~\ref{s:pp}), brief overview of the pipeline implementation (\S~\ref{s:ap}) and, description of the grid of synthetic model spectra (\S~\ref{s:sag}) and the observational data to verify the pipeline (\S~\ref{s:od}).

\subsection{Pre-processing}\label{s:pp}

The pipeline requires that all observational data are read in at the beginning as the spectra are required to be analysed all at once to determine iteratively the stellar parameters and model uncertainties (\S~\ref{s:meth_sum}). After a spectrum is loaded it is corrected for the potential radial velocity shift, transformed to the wavelength sampling of the synthetic spectra grid (\S~\ref{s:sag}) and decomposed into principal components using the decomposition matrix calculated from the synthetic spectra to reduce the memory usage and speed up the analysis. This is essential, when analysing sample of spectra of more than a few hundred sources. 

The decomposed grid of synthetic spectra is loaded into shared-memory for parallelisation purposes. The spectra are pre-convolved with combinations of varying broadening parameters of projected rotational velocity ($\varv_{\rm eq} \sin i$) and macro-turbulent velocity ($\varv_{\rm mac}$). The synthetic grid preparation is also a pre-processing step, which is laid out in  \S~\ref{s:sag}. If the sample is small and/or sufficient random access memory of the computing system is available, the grid can be convolved with the star specific broadening parameters. Even though the convolution is applied utilising the fast Fourier transformation library of {\it scipy}, this could increase the pre-processing time scale up to a few hours per star depending on the size of the synthetic spectra grid ($\gg 100\,000$).

On a standard Desktop computer 1000 spectra can be analysed in less than half a day. Larger set of samples we would advise to sort them into groups of similar objects, which will also lead to more representative model error at a specific parameter space when testing implemented physics or verifying assumptions in the theoretical model.  

\subsection{Analysis pipeline}\label{s:ap}
\begin{figure*}
\begin{center}
 \includegraphics[width=\columnwidth]{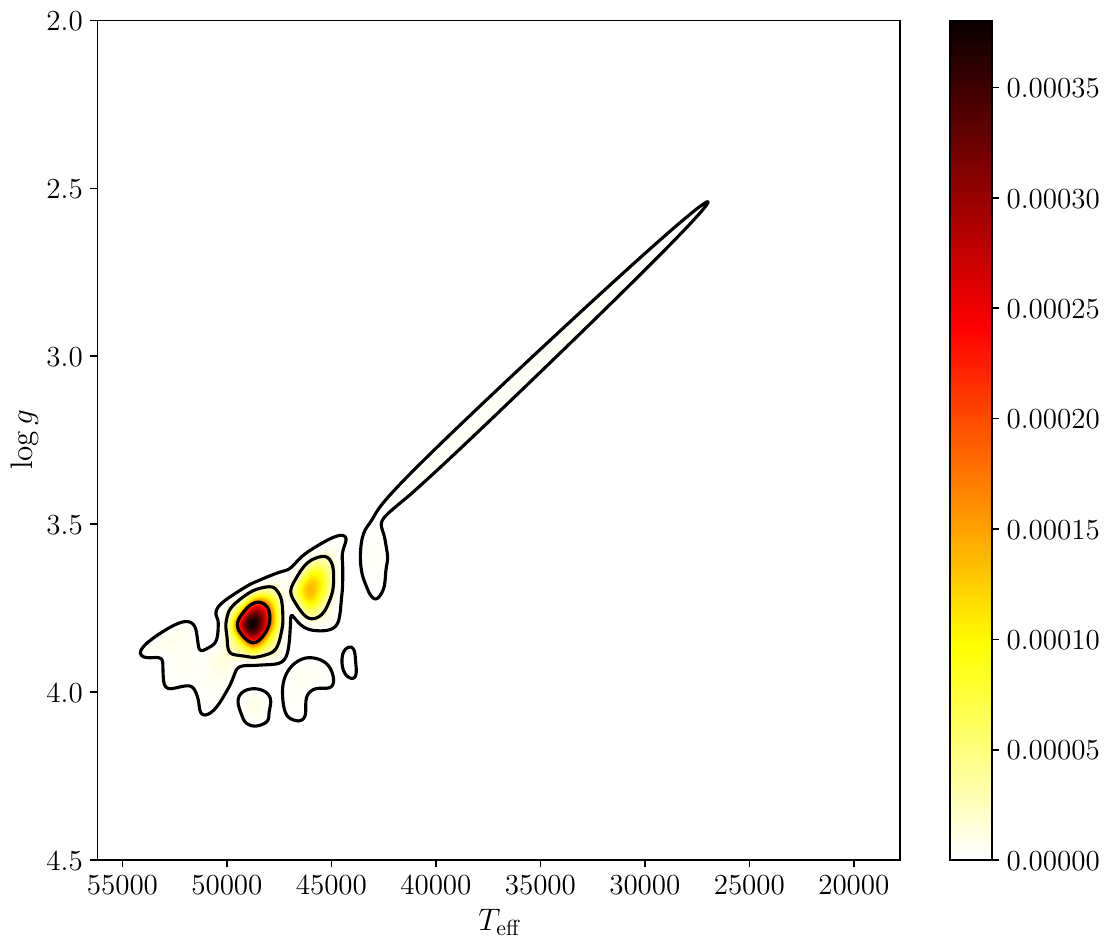} 
 \includegraphics[width=\columnwidth]{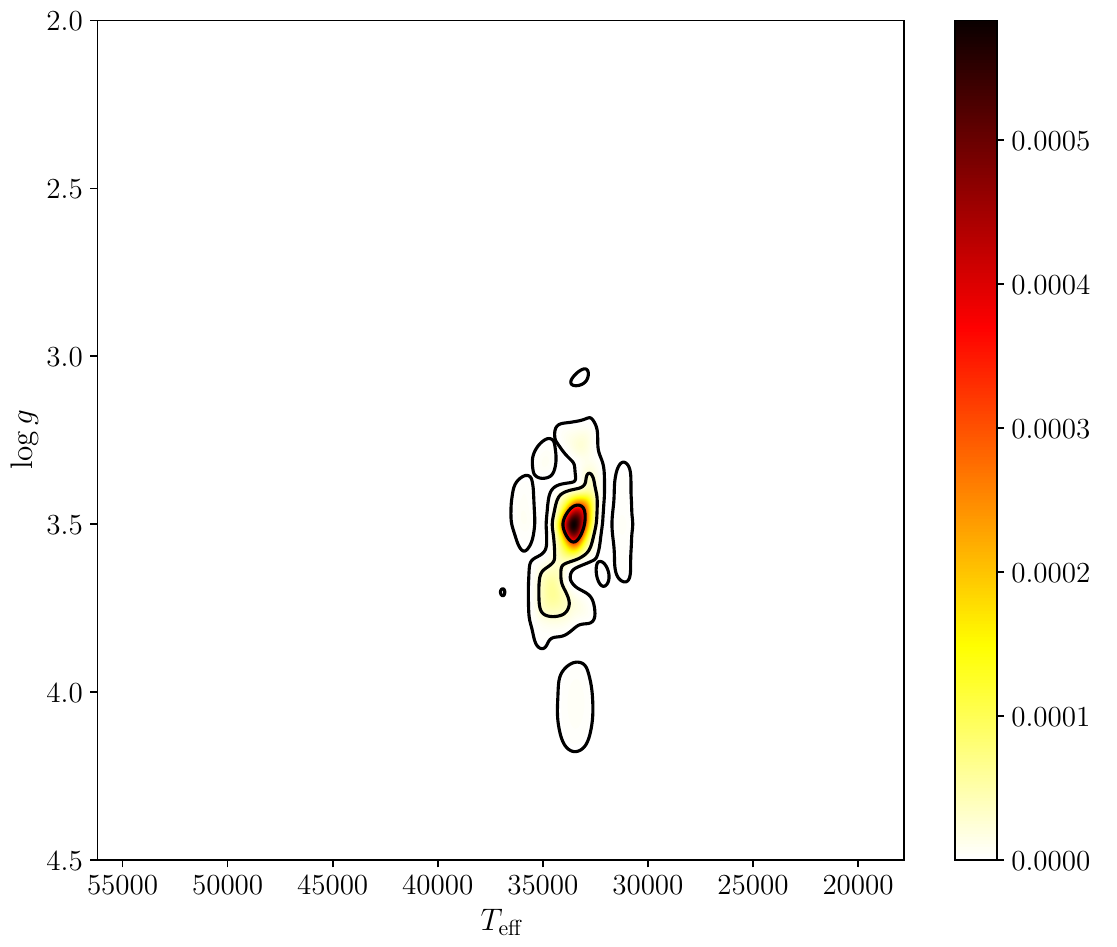}
 \caption{Probability heat map of surface gravity vs. effective temperature for VFTS-072 (left) and VFTS-076 (right). Contours indicate 2D standard-deviational ellipse confidence-intervals of 39.4\%, 86.5\% and 98.9\% \citep[e.g.][]{2015PLoSO..1018537W}. Spectroscopic fits of those stars are shown in Fig.~\ref{f:vfts_ex}.}
   \label{f:conf}
\end{center}
\end{figure*}

After pre-processing the observational data including observational error spectra and loading the synthetic grid of model spectra the pipeline is set up according to Equation~\eqref{eq:measurement model} with model de-idealisation Equation~\eqref{eq:mdi} assuming a Gaussian noise model for the observational error \eqref{eq:gnm}. We are interested in the posterior distribution of the signal given the data ($\mathcal{P}(s|d)$) and apply the Bayes theorem
\begin{eqnarray}
\mathcal{P}(s|d) =\frac{\mathcal{P}(d|s)\mathcal{P}(s)}{\mathcal{P}(d)}
\end{eqnarray}
with likelihood $\mathcal{P}(d|s)$, prior $\mathcal{P}(s)$ and evidence $\mathcal{P}(d)$ to use, first, the likelihood $\mathcal{P}(d|s)$ and, second, apply the Wiener filter (\S~\ref{s:ih}) to reconstruct the signal ($P(d|s) \to P(s|d)$). We modified the likelihood as described in \S~\ref{s:ih} ($\mathcal{P}(d|s) \Rightarrow \mathcal{P}(d|p)$), probability of the data $d$ given the stellar parameters $p$.

The best model is determined from a $\chi^2$-minimisation Ansatz with model error variance matrix $U$ (Equation~\ref{eq:effektive measurement model}) to maximise the modified likelihood $\mathcal{P}(d|p)$:
\begin{eqnarray}
\chi^2 = \left(d^{(i)} - \mathrm{R^{(i)}} s^{(i)}\right)^{\mathrm T}\mathrm{N^{(i)}}^{-1}\left(d^{(i)} - \mathrm{R^{(i)}} s^{(i)}\right).
\end{eqnarray}
For the mean and model error variance matrix $v$ and $U$, we find that it is better to set $v=0$ and multiply $U$ by a factor $\alpha = [10^{-5}, 0.35, 0.7, 1.0, 1.0]$, which increases after each iteration, to avoid that bad spectroscopic fits have significant impact on $v$ and $U$ and therefore the determination of the best model. $v$ could be also set equal to non-stellar features such as telluric bands, diffuse interstellar bands and prominent interstellar lines like Ca H and K. However, interstellar contribution can significantly vary with the line of side while telluric bands change with atmospheric conditions and airmass. In cases, where the none stellar features vary on a star by star basis, those contribution can be combined with the observational error to give less weight to those spectral regions.

$\omega_{ii'}$ (Equation~\ref{eq:generalizedcritical filter}) can contain any prior information $\mathcal{P}(p)$, e.g. parameter space of similar objects, stellar structure models or population synthesis predictions. In the current implementation we use a flat prior ($\omega_{ii'} = 1$).

The pipeline returns a multi-dimensional posterior distribution function for each star while $U$ is the same for all sources analysed in one batch. Parameters and their uncertainties are determine by defining confidence intervals (Fig.~\ref{f:conf}). To increase the accuracy the posterior distribution function can be multiplied by an appropriate prior. More details on the implementation can be found in the source code of the pipeline. 

\subsection{Stellar atmosphere grid}\label{s:sag}
The grid of synthetic model spectra was computed with the non-LTE stellar atmosphere and radiative transfer code \textsc{FASTWIND} v10.6  \citep{santolaya1997, puls2005, rivero2012a} including H, He, C, N, O and Si as explicit elements. The \textsc{FASTWIND} LINES-list and FORMAL\_INPUT file is well tested and verified in the wavelength range from 4000 to 7000~{\AA}. In the FORMAL\_INPUT file we included \ion{H}{}, \ion{He}{i-ii}, \ion{C}{ii-iv}, \ion{N}{ii-v}, \ion{O}{ii-v} and \ion{Si}{ii-iv} in the wavelength range from $\lambda$3500 to 10000. On the basis of the Vienna Atomic Line Database database\footnote{e.g.~\url{http://vald.astro.uu.se/~vald/php/vald.php}} (VALD) version 3 and NIST database\footnote{\url{https://www.nist.gov/pml/atomic-spectra-database}} we added the following lines to the FASTWIND LINES-list: \ion{C}{ii} $\lambda$6784, \ion{C}{iii} $\lambda$7703 and $\lambda$9701-05-16, \ion{C}{iv} $\lambda$4647 and $\lambda$6592-93, \ion{N}{iii} $\lambda$5321-27-52, $\lambda$3935-39 and $\lambda$9402-24, \ion{N}{iv} $\lambda$3748, $\lambda$5737, $\lambda$5776-85, $\lambda$6212-15-29, $\lambda$7103-09-11-23-27-29, $\lambda$7425 and $\lambda$9182-223, \ion{O}{iii} $\lambda$3703, $\lambda$3707-15, $\lambda$3755-57-60-74-91, \ion{O}{iv} $\lambda$3560-63, $\lambda$3729-37, $\lambda$7032-54, $\lambda$5769 and $\lambda$9454-88-92, \ion{O}{v} $\lambda$5114 and $\lambda$6500, and \ion{Si}{ii} $\lambda$9413. A full list of included spectral lines can be found in the Appendix~\ref{at:lines}.

Some lines are located in the region of telluric bands, but could be of great value, if a careful telluric correction has been performed. With data from 4MOST \citep{deJong2019} and in particular 4MOST/1001MC \citep{cioni2019} we are going to verify the \textsc{FASTWIND} LINES-list beyond the well tested wavelength range utilising the pipeline of this study.

\begin{figure}
\begin{center}
 \includegraphics[width=\columnwidth]{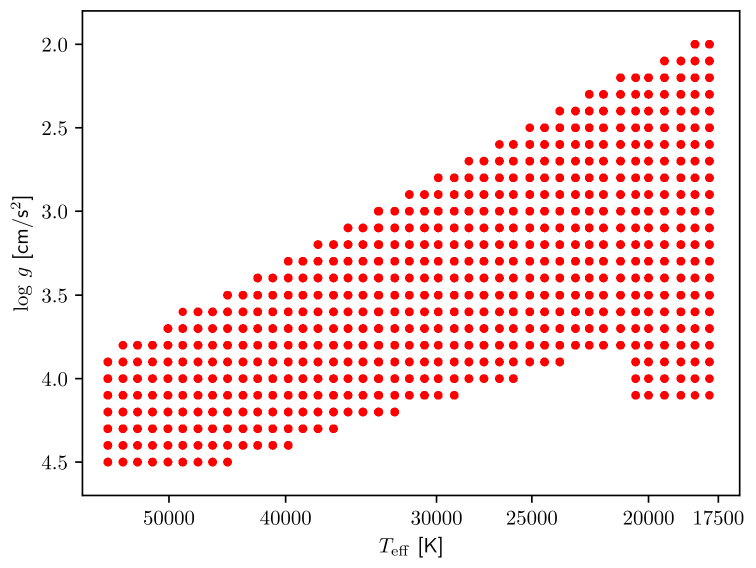} 
 \caption{$T_{\rm eff}-\log g$ plane of the computed grid of converged stellar atmospheres. The high temperature low surface gravity regime (upper left area) is empty as those models are unstable due to the Eddington limit ($\Gamma_{\rm e} \approx 0.7$). Low temperature and high surface gravity models would be better calculated with a plane-parallel code without stellar winds.}
   \label{f:kiel}
\end{center}
\end{figure}

The grid covers the parameter space for the effective temperature from $T_{\rm eff} = 17\,800$~K ($\log T_{\rm eff} = 4.25$) to 56\,200~K ($\log T_{\rm eff} = 4.75$), surface gravity $\log g/(\mathrm{g~cm^{-2}}) = 2.5$ to 4.5, transformed mass-loss rate \citep[e.g.][]{bestenlehner2014} from $\log \dot{M}_{\rm t}/(\mathrm{M_{\odot}~yr^{-1}}) = -6.5$ to $-5.0$ assuming a constant radius and helium abundances by number from $Y=0.07$ to 0.15. 3 combination of CNO abundances representing LMC baseline abundances plus semi and fully-processed CNO composition due to the CNO-cycle according to 60\,$M_{\odot}$ evolutionary track by \citet{Brott2011}. Figure~\ref{f:kiel} shows the parameter space of the grid with respect to $\log g$ and $T_{\rm eff}$. 

The high temperature, low surface gravity regime (upper left area) is unpopulated as those models are unstable as they exceed the Eddington limit at an Eddington parameter $\Gamma_{\rm e} \approx 0.7$ considering only the electron opacity $\chi_{\rm e}$. Low temperature and high surface gravity models can be computed with FASTWIND, $T_{\rm eff}$ between 17\,800~K ($\log T_{\rm eff} = 4.25$) and 21\,400~K ($\log T_{\rm eff} = 4.33$) and $\log g > 4.0$, but a significantly larger number of depth points or high mass-loss rates would be required to make them converge. The computational timescale exceeds 1 day in contrast to less than an hour. However, enhanced mass-loss rates are only observed, if the star is to the proximity of the Eddington limit \citep{bestenlehner2014}. Therefore, those low temperature and high surface gravity stellar atmosphere models are better calculated with a plane-parallel geometry code without stellar winds \citep[e.g. TLUSTY][]{1995ApJ...439..875H, 2007ApJS..169...83L}.

The clumping factor was set to $f_\textrm{cl} = 1$, i.e. a homogeneous stellar wind is adopted. We assumed a wind acceleration parameter of $\beta = 1.0$ and a fixed micro turbulence velocity of $10$\,km/s. The terminal velocity was calculated based on $\log g$ and stellar radius of the model using the escape-terminal velocity relation of $\varv_{\rm esc}/\varv_{\infty} = 2.6$ for models hotter than 21\,000~K and $\varv_{\rm esc}/\varv_{\infty} = 1.3$ for cooler models \citep{lamers1995}. In total we computed of the order $\lesssim 150\,000$ stellar atmospheres. For around $\sim 20\%$ of those models the atmospheric structure, ionisation balance and/or radiation field failed to converge properly leading to negative fluxes, discontinuities or even failed when the spectral lines were synthesised. 

The grid was then convolved with a macro-turbulent velocity of $\varv_{\rm mac} = 20$\,km/s and varying projected rotational velocity $\varv \sin i = [0, 20, 50, 100, 150, 200, 250, 300, 400]$\,km/s assuming $\varv_{\rm eq} \sin i$ is the dominant broadening mechanism, which is a reasonable assumption given the spectral resolution of the observational data (\S~\ref{s:od}) and that typical $\varv_{\rm mac}$ are of the order of a few 10s~km/s. This results in a grid of $\lesssim 1\,100\,000$ synthetic spectra, which has been used to compute the decomposition matrix to decomposed the grid into its principal components reducing the size by a factor $\sim 200$. The decomposition matrix is also used to decompose the observational data ({\S~\ref{s:pp}).

\subsection{Observational data}\label{s:od}
To validate the methodology of the pipeline we used the VLT/FLAMES data of VLT/FLAMES Tarantula survey \citep[VFTS][]{evans2011} in the traditional blue-optical wavelength regime with the LR02 ($\lambda3964-4567$, $\lambda/\Delta\lambda=6000$}), LR03 ($\lambda4501-5078$, $\lambda/\Delta\lambda=7500$) and HR15N ($\lambda6470-6790$, $\lambda/\Delta\lambda=19\,000$) gratings. We selected 240 O-type stars employing the same data and normalisation as used in \citet{bestenlehner2014, sabin2014, sabin2017, ramirez2017} to avoid the introduction of biases.  

The second data set we used are the VLT/MUSE observation of $\sim 250$ OB stars from \citet{castro2018}. The VLT/MUSE data cover the wavelength range between 4600 to 9300 {\AA} at spectral resolution of 2000 to 4000. The normalisation of the spectra was fully automated simulating the work-flow of the spectroscopic analysis of large datasets. 35 stars are in common with VFTS, which will allow us to test the reliability of line diagnostics towards the red of $\mathrm{H}_{\alpha}$ (\S~\ref{s:sag}) and the automated normalisation routine.

\section{Results and discussion}\label{s:re_di}
\subsection{VLT/FLAMES: \citet{evans2011}\label{s:vfts}}
\begin{figure*}
\begin{center}
 \includegraphics[width=\columnwidth]{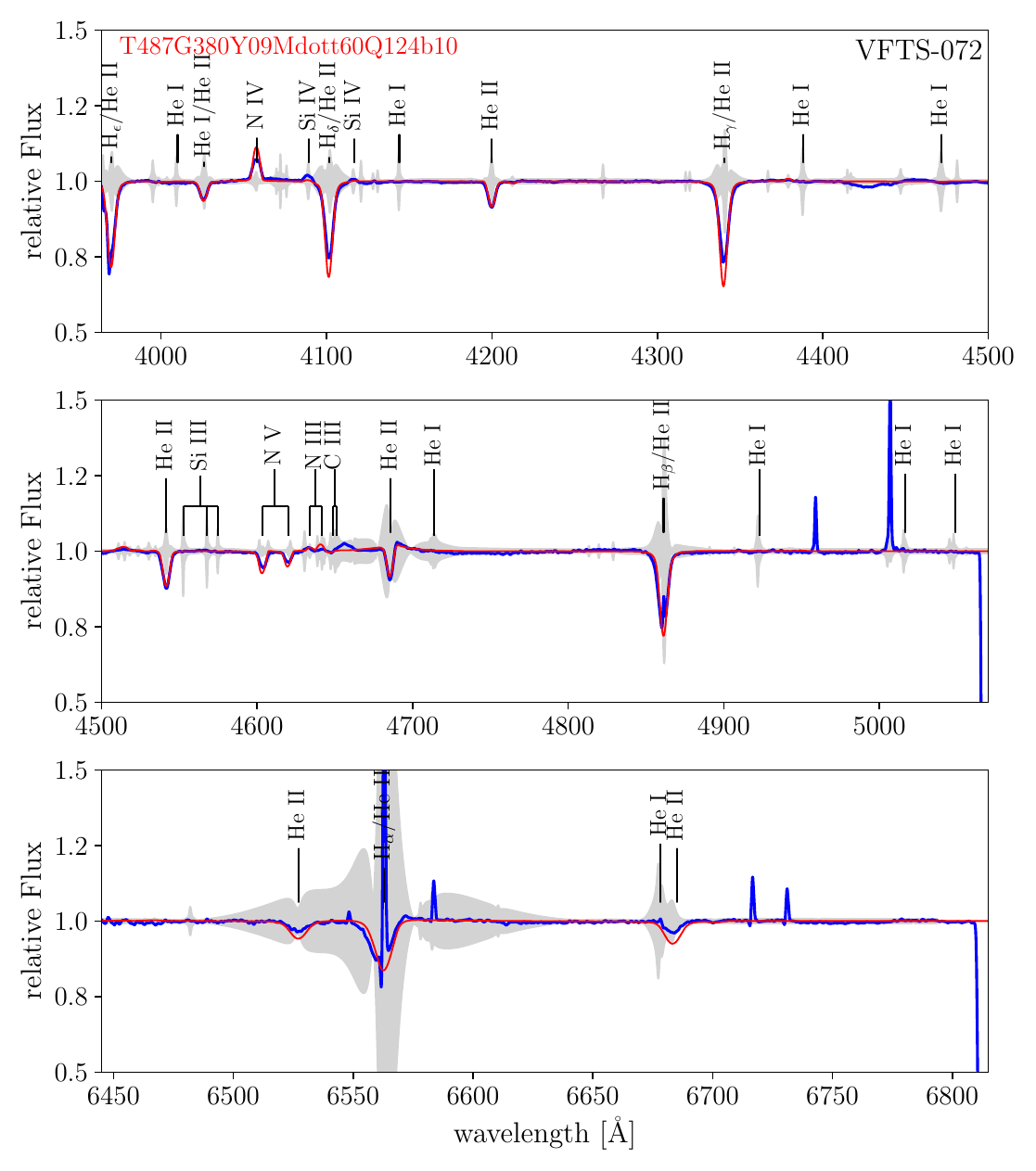} 
 \includegraphics[width=\columnwidth]{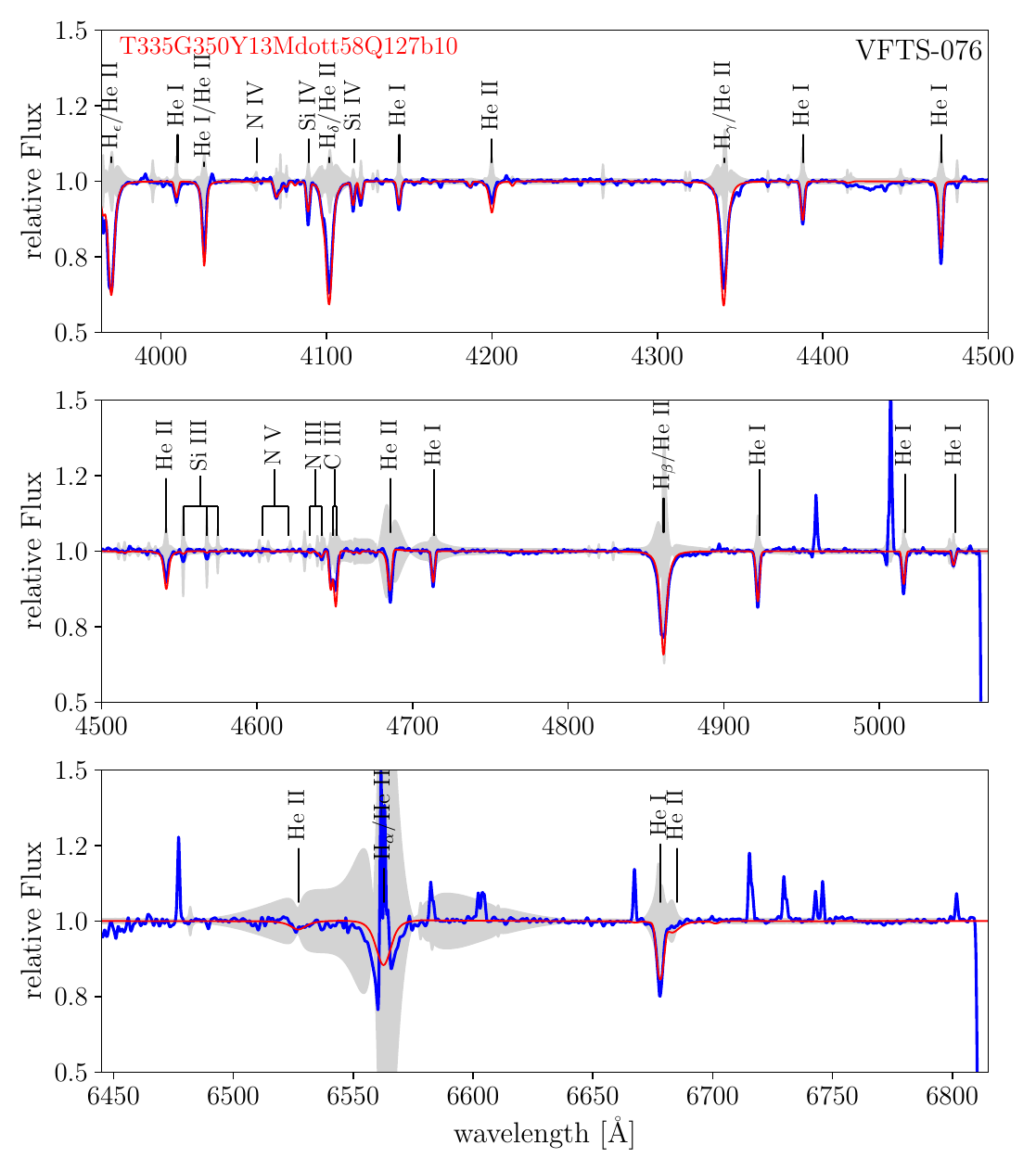}
 \caption{Left, spectroscopic fit of an fast rotating early O2 V-III(n)((f*)) star VFTS-072 and, right, a late O9.2 III star VFTS-076 (right). Blue solid line is the observation, red solid line the synthetic spectrum and the grey shaded area is the square-root of the diagonal elements of the model-error uncertainty-matrix calculated by the pipeline.}
   \label{f:vfts_ex}
\end{center}
\end{figure*}
We analysed all 240 O-type stars including stars with low SNR and/or strong nebular contamination (e.g. Fig.
\ref{af:142}) while \citet{bestenlehner2014, sabin2014, sabin2017, ramirez2017} provided reliable results for 173 out of 240 sources. In Fig.~\ref{f:vfts_ex} we show the spectroscopic fits of a representative early and late O star (VFTS-072 and 076). The shaded error area in Fig.~\ref{f:vfts_ex} reveal where a general mismatch between model and observations occurs and is the square-root of the diagonal elements of the model-error uncertainty-matrix. In particular the line centres of the Balmer and \ion{He}{i-ii} lines seemed to be poorly fitted as a results of nebular lines, inaccurate determined line-broadening, insufficient grid resolution and range for helium abundances, fixed micro-turbulent velocity or shape of line profiles. We are also able to locate spectra lines, which are potentially not included into the \textsc{FASTWIND} LINES-list or require improved atomic data (e.g. \ion{Si}{iii} $\lambda$4552.6, 4567.8 and 4574.7). Overall the spectroscopic fit is good for the synthetic spectra (red solid line) to the observations (blue solid line).

Figure~\ref{f:conf} visualises the probabilities in the $\log g - T_{\rm eff}$ plane. VFTS-072 (left panel) shows within 2-sigma a dual solution ($\sim$49\,000 and $\sim$46\,000~K). At $\sim 45\,000$~K the \ion{He}{i} disappear, but the \ion{N}{iv} and {\sc{v}} lines sufficiently contributed to the $\chi^2$ so that the correct solutions around $49\,000$~K has also the highest probabilities. By looking at the 3-sigma contour we notice a degeneracy between $\log g - T_{\rm eff}$ due to the proximity of VFTS-072 to the Eddington limit. In contrast, the heat map of VFTS-076 (right panel) is well centred on a specific $\log g - T_{\rm eff}$ region. However, a slightly higher surface gravity could be probable within 2-sigma, which is the results of a degeneracy between surface gravity and mass-loss rates. High mass-loss rates fill in the wings of the Balmer lines mimicking a lower surface gravity.

\subsubsection{Model-error uncertainty-matrix}
\begin{figure}
\begin{center}
 \includegraphics[width=\columnwidth]{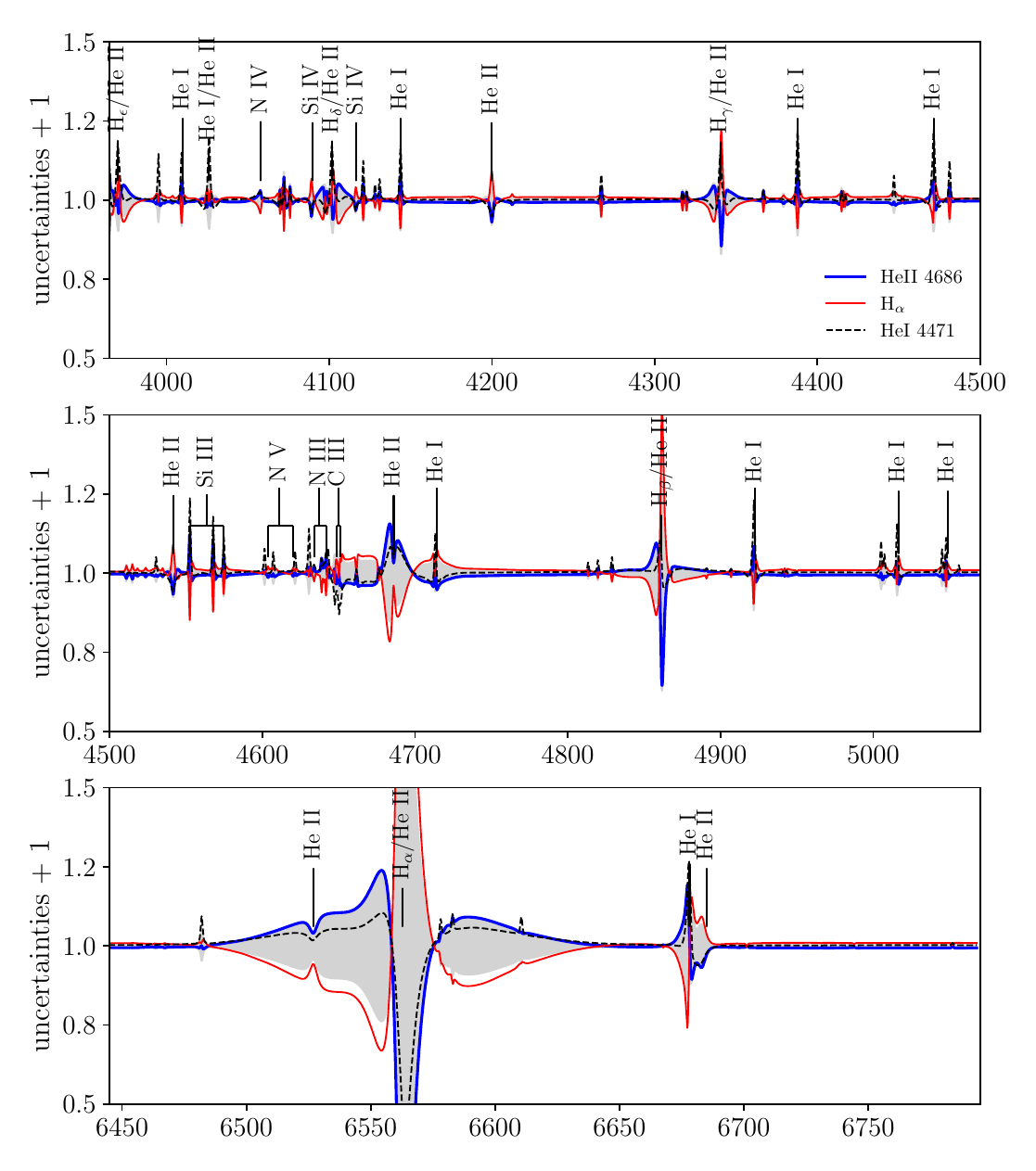} 
 \caption{Model uncertainties extracted from model-error matrix as a function of wavelength for $\mathrm{H}_{\alpha}$ (solid red), \ion{He}{i} $\lambda4471$ (black dashed) and \ion{He}{ii} $\lambda4686$ (solid blue). For better visualisation uncertainties for \ion{He}{i} $\lambda4471$ and \ion{He}{ii} $\lambda4686$ are multiplied a factor 25 and 10, respectively.}
   \label{f:mod_err_ex}
\end{center}
\end{figure}

The model-error uncertainty-matrix is symmetric ($U^{\rm T} = U$) and shows correlation between wavelength or pixel regions. An example is given in the Appendix (Table~\ref{at:err_mat}), where we reduced the rank of the matrix from 11840$\times$11840 to 37$\times$37 for visualisation purposes. The strongest correlations are between the Balmer lines, which are the most prominent lines in O-type stars. On the other hand \ion{He}{i} is present in mid to late O stars, while \ion{He}{ii} lines are only strong in early O stars. Therefore, to visualise the model-error matrix and its correlations we plot in Fig.~\ref{f:mod_err_ex} the model uncertainties for wavelengths of $\mathrm{H}_{\alpha}$, \ion{He}{i} $\lambda4471$ and \ion{He}{ii} $\lambda4686$.    

$\mathrm{H}_{\alpha}$ and \ion{He}{ii} $\lambda4686$ are anti-correlated with each other, although we amplified the uncertainties for \ion{He}{ii} $\lambda4686$ by a factor of 10. Wavelength regions of Balmer lines are a blend of hydrogen and \ion{He}{ii} lines. With increasing temperature the \ion{He}{ii} lines are stronger while the Balmer lines become weaker. In addition, the line strength between hydrogen and helium also determines their abundances, e.g. overall stronger helium lines with respect to hydrogen lines mean lower hydrogen abundances. 

\ion{He}{i} $\lambda4471$ (amplified by a factor of 25) is correlated with \ion{He}{ii} $\lambda4686$ for helium lines, but  anti-correlated for $\mathrm{H}_{\delta}$ and higher order Balmer lines following the trend of $\mathrm{H}_{\alpha}$. \ion{He}{i} $\lambda4471$ show stronger correlations with \ion{Si}{iii} and \ion{C}{iii}, which are only present in late O stars, where \ion{He}{i} lines are strongest as well. Under this supposition we would expect that we observe a strong correlation between \ion{He}{ii} and the higher ionised \ion{N}{iv} and {\sc{v}}, which seems not to be the case. A reason might be that the number of early O stars is too small due to the stellar mass-function to significantly contribute to the model-error ($<5$\%). Grouping similar objects together is advisable when testing model assumptions in stellar atmosphere codes.

\subsubsection{Challenges}

The examples shown in Fig.~\ref{f:vfts_ex} show low and modest nebular contamination and the pipeline derives results in good agreement with VFTS. However, the pipeline has difficulties when spectra show strong nebular lines. In the case of VFTS-142 (Fig.~\ref{af:142}) the temperature is still reasonably well reproduced, but the surface gravity is by $\sim$0.3~dex too low. If the spectra is dominated by nebular lines, the pipelines will fail, e.g. VFTS~410 (Fig.~\ref{af:410}). Often nebular lines are clipped, but in the case of VFTS-410 only few diagnostic lines would remain. Even though we perform a single star analysis, for double-lined spectroscopic binaries (SB2s) the pipeline is able to fairly fit the primary component, but struggles with the mass-loss rate and helium abundances due to the contribution of the colliding wind region of VFTS-527 \citep[Fig.~\ref{af:527},][]{taylor2011}.

\begin{figure*}
\begin{center}
 \includegraphics[width=\columnwidth]{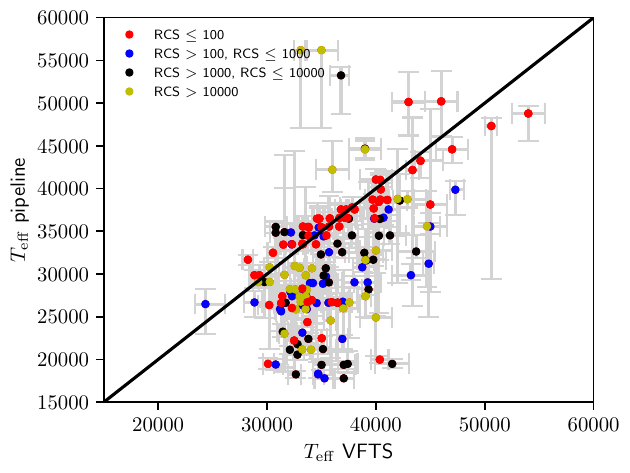} 
 \includegraphics[width=\columnwidth]{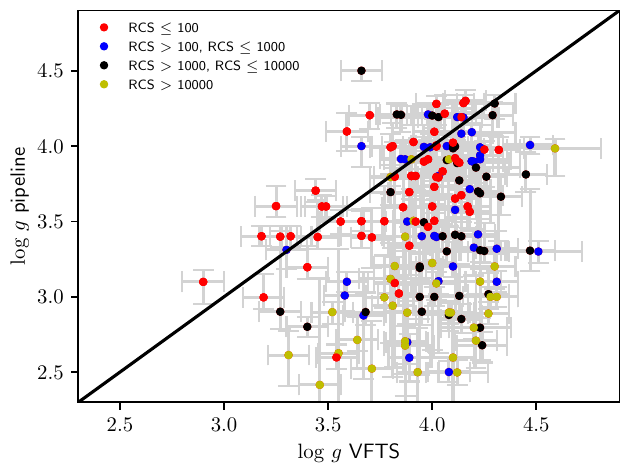} 
 \caption{Effective temperatures (left) and surface gravities (right) determined by the pipeline vs. the results from \citet{bestenlehner2014, sabin2014, sabin2017, ramirez2017}.}
   \label{f:vfts_Tg}
\end{center}
\end{figure*}
The goodness-of-fit is usually evaluated by calculating the reduced-chi-square (RCS), which uses in our case the diagonal of the error co-variance matrix. Due to the nebular contamination and diffuse interstellar bands none of our fits had a RCS close to 1. To visualise how well the pipeline performs we compare our results versus tailored analysis of VFTS targets in Fig.~\ref{f:vfts_Tg}. Our results agree well with \citet{bestenlehner2014, sabin2014, sabin2017, ramirez2017} for fits with RCS $<100$. Effective temperatures show a tighter relation than the surface gravity. The determination of surface gravity is based on the wings of the Balmer lines, which is influenced by the line broadening and therefore how well $\varv_{\rm mac}$ and $\varv \sin i$ are determined. If the spectroscopic fit is poor, we derive systematically lower temperatures and surface gravities. Low temperature and gravity models have $\mathrm{H}_{\alpha}$ in emission to somehow fit the none-stellar $\mathrm{H}_{\alpha}$ nebular line while higher order Balmer lines remain in absorption. Overall there is good agreement considering that our analysis took less than 30~min while the VFTS analysis involved 3 PhD theses.

Looking at the error bars the pipeline obtains systematically larger uncertainties in part due to the inclusion of the model error but mostly as a result of the interpretation of the 4D posterior distribution function (PDF), which includes the degeneracies between $T_{\rm eff}$, $\log g$, $\dot{M}_{\rm t}$ and $Y$. The derived errors might be larger, but they are a complete representation of the true uncertainties. A representative prior could increase the accuracy, but might introduce additional biases \citep{bestenlehner2020b}. Temperature uncertainties are systematically larger in the region around 45,000~K as a result of the weakness of \ion{He}{i} lines and the ionisation balance is not based on \ion{He}{i-ii}, but on the metal ions \ion{N}{iii-iv-v}. Nitrogen lines in early O star are relatively weak compared to He lines and therefore contributed little to the global $\chi^2$ without specific weighting of spectral lines. This can lead to an overall lower RCS, but inaccurate temperature determination (red outliers in Fig.\ref{f:vfts_Tg}). A similar behaviours occurs in the transition from late O to early B stars due to the weakness of \ion{He}{ii} lines, where the main temperature diagnostics are \ion{Si}{iii} and {\sc iv} lines. So a careful weighting scheme should be a very promising way for optimising the pipeline by increasing the accuracy while at the same time reducing the degeneracies between parameters.

\subsection{VLT/MUSE: \citet{castro2018}}
\begin{figure*}
\begin{center}
 \includegraphics[width=\columnwidth]{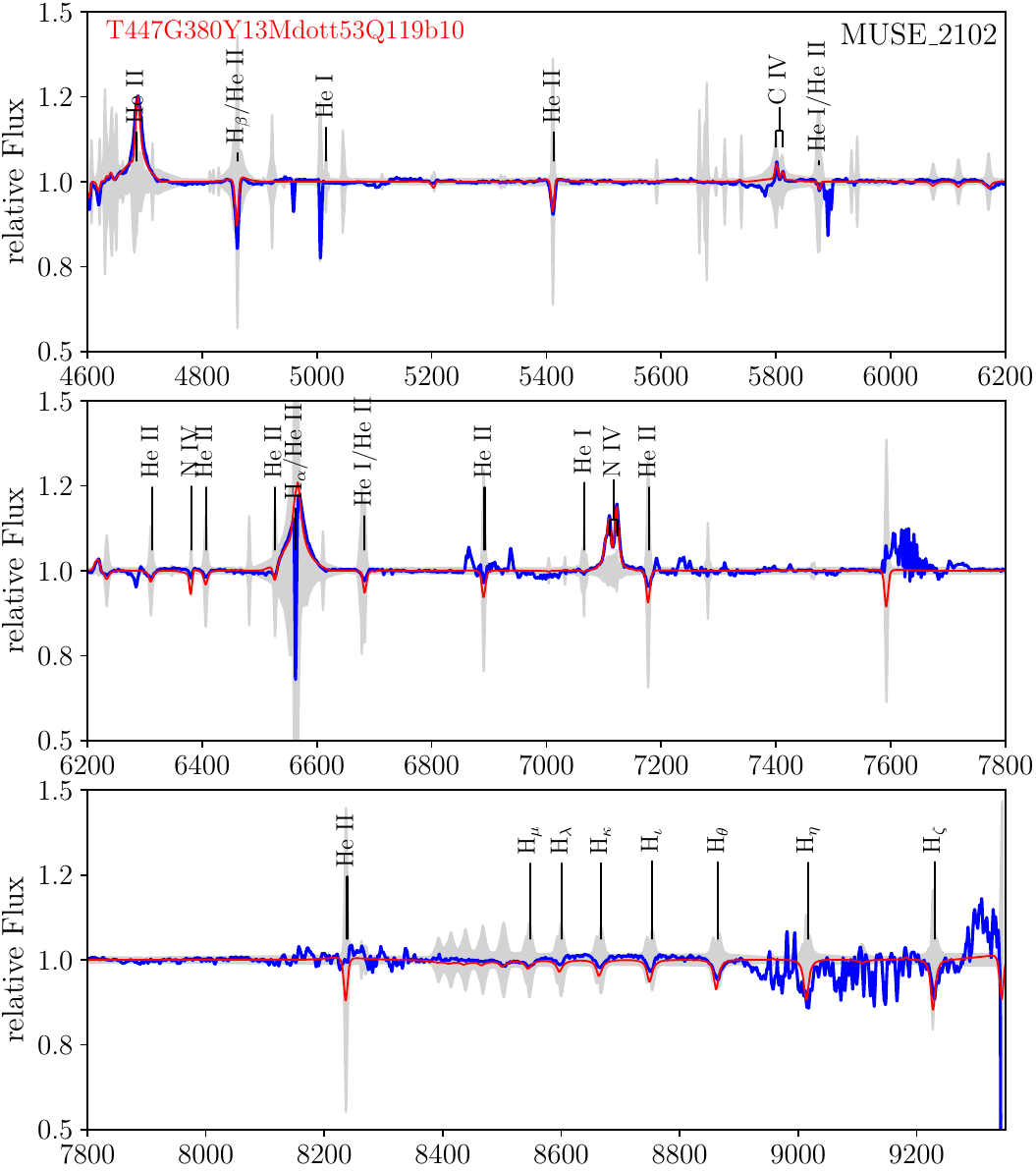} 
 \includegraphics[width=\columnwidth]{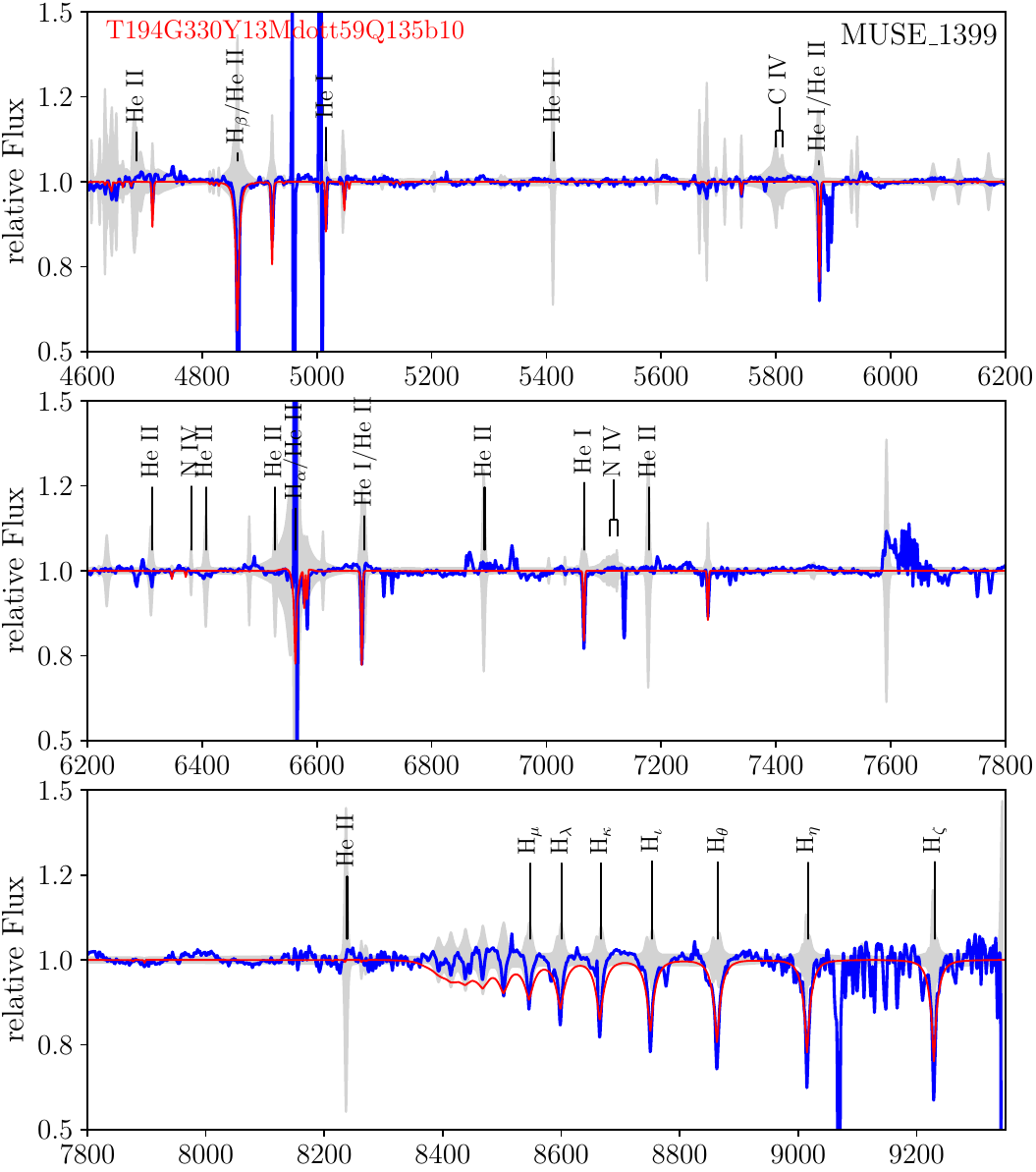}
 \caption{Spectroscopic fit of an O2\,If (Mk42, left) and a B2\,Ib (VFTS-417, right). Blue solid line is the observation, red solid line the synthetic spectrum and the grey shaded area is the square-root of the diagonal elements of the covariant-matrix calculated by the pipeline. In the left panel the newly added \ion{N}{iv} mulitplet at $\lambda7103-29$ is able to reproduce the observed line.}
   \label{fig2}
\end{center}
\end{figure*}
Figure~\ref{fig2} shows the spectroscopic fit of an Of supergiant and a B supergiant with $\Delta T_{\rm eff} \approx 25\,000$\,K and $\Delta \log\dot{M} \gtrsim 2.5$\,dex. This highlights that stars covering a large spectral type range can be successfully and reliably analysed with a single pipeline set up at the same time. However, both stars would not be considered as similar, which has implication on the model error $U$. Such a model error is averaged over a wide parameter space and is probably not very helpful, when testing specific physics (e.g. stellar wind physics or atomic data) in the model. Similar to the VFTS data the pipeline performs not well for low signal to noise spectra (S/N $\lesssim 10$ to 15), spectra with strong nebular lines and spectroscopic binaries/multiples.

\begin{figure*}
\begin{center}
 \includegraphics[width=\columnwidth]{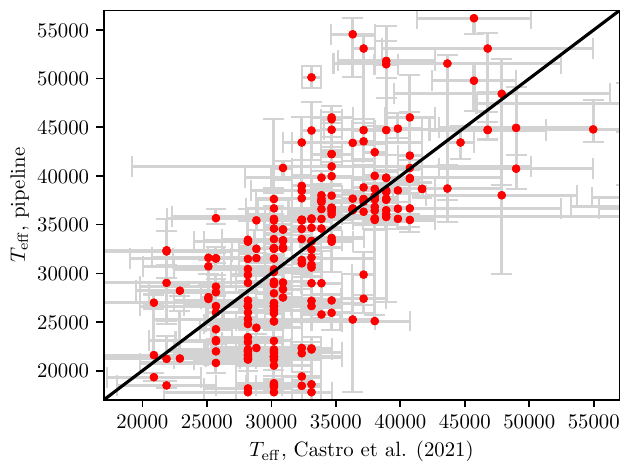} 
 \includegraphics[width=\columnwidth]{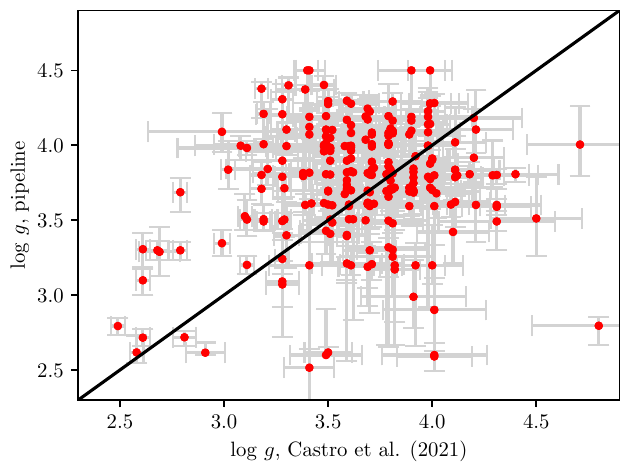} 
 \caption{Effective temperatures (left) and surface gravities (right) determined by the pipeline vs. the results from \citet{castro2021}.}
   \label{fig3}
\end{center}
\end{figure*}

In Fig.~\ref{fig3} we compare our results with those from \citet{castro2021} which is based on the ionisation balance of selected HeI and HeII and the wing of $H_\mathrm{\beta}$. In contrast, we used all H, He plus CNO and Si metal lines available in the VLT/MUSE wavelength range. The left panel compares effective temperatures, which shows a large scatter, but mostly agrees within their large uncertainties. Above 45\,000~K, when \ion{He}{i} disappears or weakly present in the spectra, the temperature needs to be derive based on the ionisation balance of metal lines. In the wavelength range of MUSE we have \ion{C}{iii-iv} and \ion{N}{iii-iv}. While \ion{C}{iv} and \ion{N}{iv} are located in relative clean areas of the spectra, the \ion{C}{iii} and \ion{N}{iii} are often found in the range of telluric bands or near the Paschen jump, where we have issues with the normalisation (Fig.~\ref{fig2}). B type stars have per definition no \ion{He}{ii} present in their spectra and the temperature is based in the case of early B stars on the ionisation balance of \ion{Si}{iii-iv} lines. There is a reasonable number of lines in the MUSE range, but the temperature determination suffers with the presence of nebular lines or low SNR spectra.  

The right panel of Fig.~\ref{fig3} compares surface gravities which show an even larger scatter and uncertainties. Even though we utilised the Paschen lines as well, there are 2 potential caveats, first, the normalisation near the Paschen jump is not straightforward as the lines overlap and therefore no continuum, second, the degree of overlapping depends not only on $\log g$ but also on the line broadening due to the narrowness of the higher order Paschen lines, which is only approximately determined during the fitting process. This results in a degeneracy between $\log g$ and $\varv_{\rm eq} \sin i$. Surface gravities cluster in the range of $\log g = 3.5$ to 4.5, which is expected for a young stellar population of 1 to 2 Myr largely consisting of dwarfs and giants in the proximity of R136 \citep{bestenlehner2020b}.

\begin{figure*}
\begin{center}
 \includegraphics[width=\columnwidth]{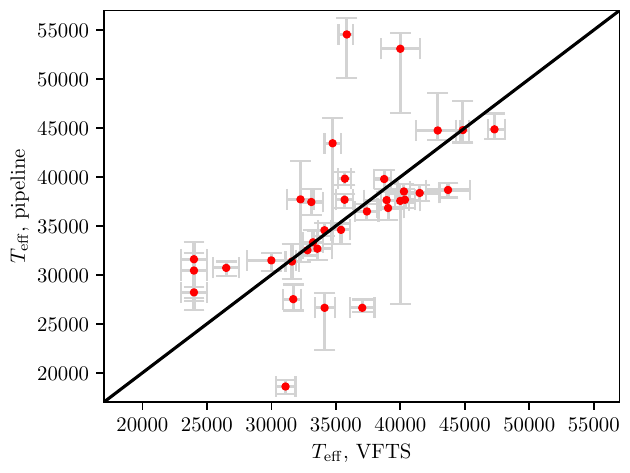} 
 \includegraphics[width=\columnwidth]{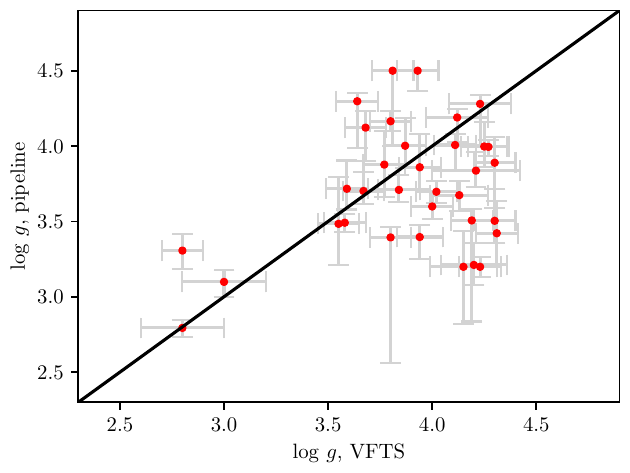} 
 \caption{Comparison of MUSE targets in common with VFTS: effective temperatures (left) and surface gravities (right) determined by the pipeline using VLT/MUSE data vs. the results from \citet{bestenlehner2014, sabin2014, mcevoy2015, sabin2017, ramirez2017} using VLT/FLAMES data.}
   \label{f:muse_vfts}
\end{center}
\end{figure*}

To better quantify how reliable the analysis based on the VLT/MUSE data is, we compare our results to VFTS. 35 VLT/MUSE targets are in common with VFTS \citep{bestenlehner2014, sabin2014, sabin2017, ramirez2017} and the comparison is shown in Fig.~\ref{f:muse_vfts}. Uncertainties are systematically larger. Effective temperatures agree within their 1$\sigma$ uncertainties for 26 out 35 stars (left panel) with differences largely as a result of the cleanliness and quality of spectra. Surface gravities are in consent for only half of the sample, which is expected due to the challenges of the Paschen lines. Overall the agreement is reasonable considering the difficulties in the analysing of the MUSE data.  

\begin{figure}
\begin{center}
 \includegraphics[width=\columnwidth]{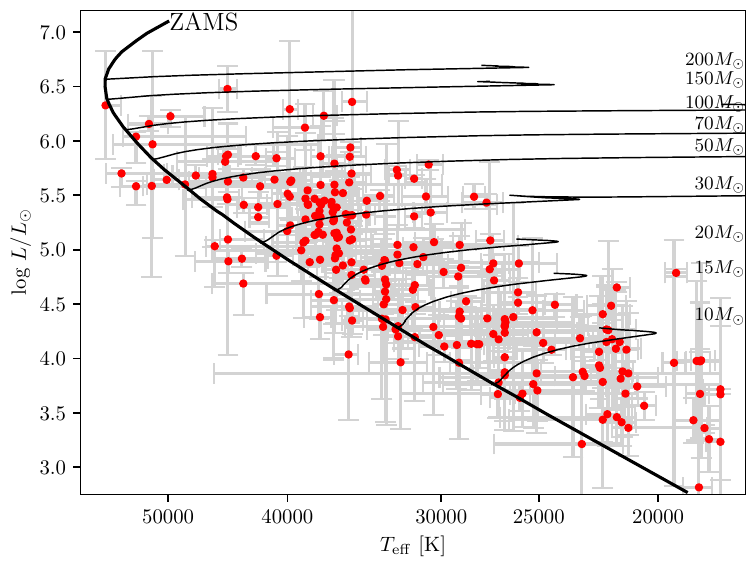} 
 \caption{Hertzsprung-Russell diagram of the analysed stars using the VLT/MUSE data from \citet{castro2018}. Thin black lines are stellar evolutionary tracks by \citet{Brott2011} and \citet{koehler2015}.}
   \label{fig4}
\end{center}
\end{figure}

To place the stars into the Hertzprung-Russell diagram (HRD) we derived bolometric luminosity on the basis of optical $UBV$ from \citet{selman1999} and near-infrared $JK_{\rm s}$ photometry from 2MASS ((2 Micron All Sky Survey)) and Vista Magellanic Clouds survey \citep[VMC][]{cioni2011} using the methodology of \citet{bestenlehner2020b, bestenlehner2022}. The HRD (Fig.~\ref{fig4}) shows that most stars are populated near, and to the cool side of the zero-age main-sequence (ZAMS). There are a couple of exceptions but their uncertainties do not exclude a cooler location in agreement with the majority of the sources. This can be improved by including a meaningful prior into the analysis, e.g. based on evolutionary tracks, and could increase the accuracy of the results, as we used only a flat prior ($\omega_{ii'} = 1$). For example, only hydrogen deficient stars are found to be on the hot side of the ZAMS, e.g. self-stripping through stellar winds or binary evolution. A prior would give the star a higher probability to be found either on the hot or cool side of the ZAMS depending on its helium composition. Stellar parameters can be found in Table~\ref{at:muse_sp} while individual spectroscopic fits for visual inspection are in the supplementary online material. Mass-loss rates, and He, C, N, and O abundances are not included. Optical data can only provide an upper limit for most stars in our sample (flat PDF towards lower mass-loss rates) while the PDF for Helium is cut off at the primordial abundance of 25\% by mass. CNO abundances are too coarsely sampled and linked to the predicted chemical evolution of 60~$M_{\odot}$ star (\S~\ref{s:sag}).

\section{Conclusions and Outlook}\label{s:con_out}

Large spectroscopic surveys with WEAVE and 4MOST will observed 10\,000s of massive stars, which need to be analysed in a homogeneous and efficient way. The pipeline presented in this work takes advantage and utilises the information that large data sets provide by determining the model uncertainties, which are included into the error budget. This methodology could be also applied to galaxies or other domains like biology and geophysics for which approximate and incomplete theoretical models exist as well. 

The runtime of the pipeline scales exponentially with the number of spectra, because all stars are analysed at the time and the error-model uncertainties-matrix is iteratively updated (\S~\ref{s:meth_sum}). However, once a converged error-model uncertainties-matrix is obtained, we can limit the matrix operations to the $\chi^2$-minimisation and switch to a star by star analyses. In this case we are able to analyse 1 star in less than a second.

\citet[{\sc The HotPayne}]{2022A&A...662A..66X} applied the FGK methodology of {\sc The Payne} \citep{2019ApJ...879...69T} to OBA stars to derive 2 stellar labels/parameters ($T_{\rm eff}$ and $\log g$) plus 15 chemical abundances. They used the plane-parallel LTE atmospheric model calculated with {\sc ATLAS12} \citep{1970SAOSR.309.....K, 1993sssp.book.....K, 2005MSAIS...8...14K} and were able to analyse 330\,000 spectra. While $T_{\rm eff}$ and $\log g$ are sensible for A and mid-late B dwarfs the derived chemical abundances suffer from non-negligible systematics due to non-LTE effects. AB supergiants require spherical geometry (stellar radius $\ngg$ scale height) including stellar winds as these effects cannot be neglected. In hotter and more luminous stars (early B and O stars) a non-LTE treatment is necessary \citep{mihalas1970} such that {\sc The HotPayne} results for $T_{\rm eff}$ and $\log g$ are not reliable including stars with weak winds. Even the inclusion of a model error will not improve much due to the fundamental missing physics in the {\sc ATLAS12} models.

To make {\sc The HotPayne} usable for OBA stars the underlying stellar models must be replaced with models computed with more sophisticated and fully non-LTE stellar atmosphere codes designed for hot, massive stars with stellar winds, e.g. {\sc cmfgen} \citep{hillier1998}, {\sc fastwind} \citep{santolaya1997, puls2005} or {\sc PoWR} \citep{graefener2012, Hamann2003}. The approach of the {\sc The HotPayne} needs to be expanded: to include the model uncertainties into the error budget (e.g. this work) to account for the assumptions and parametrisations utilised in those complex stellar atmosphere codes. Additional stellar labels/parameters need to be incorporated such as mass-loss rate, velocity law, wind-inhomogeneity, terminal and wind-turbulent velocity plus Helium abundances. The helium abundance increases due to the CNO cycle, which in turn increases the mean molecular weight ($\mu$), impacts the mass-luminosity relation ($L\propto \mu^4 M^3$), electron density, and therefore the structure and ionisation balance of the stellar atmosphere. When analysing optical spectra the wind parameters can be merged into a wind strength parameter $Q$ \citep{puls1996}, transformed radius $R_{\rm t}$ \citep{schmutz1989,graefener2002,hamann2004} or transformed mass-loss rate $\dot{M}_{\rm t}$ \citep[\S~\ref{s:sag}]{bestenlehner2014}.

The fully automated spectroscopic analysis tool of this work reduces the human interaction to a minimum to cope with the amount of data. It is able to process $\sim$ 250 stars in less than half an hour ($\sim$ 6 CPU hours) delivering results comparable to \citet{bestenlehner2014, sabin2014, sabin2017, ramirez2017} over a decade. Overall the quality of the spectroscopic fits are good, but around 15\% of the stars need additional attention as a result of strong nebular contamination, low S/N, multiplicity et cetera. The pipeline performances well over a wide parameter space which is support by the spectroscopic analysis of 3 benchmarck stars by several groups within the X-Shooter and ULLYSES collaboration \citep[XShootU][]{2023arXiv230506376V} of optical VLT/XShooter data (Sander et al. in preparation).

Weights of spectral lines could increase the accuracy, but need to be adjusted depending on the parameter space that would then require human interaction. Determining weights for features (spectral lines) is a typical machine learning problem and often solved with neural networks (deep learning). However, to really take advantage of our statistical approach and optimise the pipeline we will require much larger data sets, which will be soon provided by WEAVE and 4MOST. Future advances of our pipeline will be released on the pipelines repository. 

\section*{Acknowledgements}
JMB and PAC are supported by the Science and Technology Facilities Council research grant ST/V000853/1 (PI. V. Dhillon).
MB is supported through the Lise Meitner grant from the Max Planck Society. We acknowledge support by the Collaborative Research centre SFB 881 (projects A5, A10), Heidelberg University, of the Deutsche Forschungsgemeinschaft (DFG, German Research Foundation). This project has received funding from the European Research Council (ERC) under the European Union’s Horizon 2020 research and innovation programme (Grant agreement No. 949173).


\section*{Data Availability}

The data underlying this article are available in the article and in its
supplementary material. The pipeline will be made publicly available after acceptance of this manuscript. Spectroscopic data are available via the
ESO archive facility while grids of synthetic spectra can be requested from
the lead author.



\bibliographystyle{mnras}
\bibliography{reference} 




\appendix

\section{Additional material}
\begin{table*}
	\centering
	\caption{Lines synthesised when calculation the formal integral. Wavelength ranges are multiplets with diverging central wavelengths.}
	\label{at:lines}
	\begin{tabular}{lclclclclclc} 
		\hline
		Ion & Wavelength	& Ion 	& Wavelength & Ion & Wavelength	& Ion 	& Wavelength & Ion & Wavelength	& Ion 	& Wavelength\\
			& [\AA]		&	& [\AA]      & 	   &   [\AA]	&	& [\AA]	     &     &  [\AA]	&	& [\AA]\\
		\hline
		\ion{H}{i} &$3835.4$& \ion{He}{i}  &$3888.6$& \ion{C}{ii}  &$3919.0-3920.7$& \ion{N}{ii}  &$3995.9       $& \ion{O}{ii} &$3945.0-3954.4$& \ion{Si}{ii} &$3853.7-3862.6$\\
		\ion{H}{i} &$3889.1$& \ion{He}{i}  &$3964.7$& \ion{C}{ii}  &$4267.0-4267.3$& \ion{N}{ii}  &$4447.0       $& \ion{O}{ii} &$4069.6-4075.9$& \ion{Si}{ii} &$4128.1-4130.9$\\
		\ion{H}{i} &$3970.1$& \ion{He}{i}  &$4009.3$& \ion{C}{ii}  &$4637.6-4639.1$& \ion{N}{ii}  &$4530.4       $& \ion{O}{ii} &$4317.1-4366.9$& \ion{Si}{ii} &$5041.0-5056.3$\\
		\ion{H}{i} &$4101.7$& \ion{He}{i}  &$4026.2$& \ion{C}{ii}  &$5132.9-5151.1$& \ion{N}{ii}  &$4552.5       $& \ion{O}{ii} &$4414.9-4452.4$& \ion{Si}{ii} &$5957.6-5978.9$\\
		\ion{H}{i} &$4340.5$& \ion{He}{i}  &$4120.8$& \ion{C}{ii}  &$5648.1-5662.5$& \ion{N}{ii}  &$4601.5-4643.1$& \ion{O}{ii} &$4638.9-4676.2$& \ion{Si}{ii} &$6347.1-6371.4$\\
		\ion{H}{i} &$4861.4$& \ion{He}{i}  &$4143.8$& \ion{C}{ii}  &$6151.3-6151.5$& \ion{N}{ii}  &$5005.2       $& \ion{O}{ii} &$4890.9-4906.8$& \ion{Si}{ii} &$9412.7-9412.8$\\
		\ion{H}{i} &$6562.8$& \ion{He}{i}  &$4387.9$& \ion{C}{ii}  &$6461.9       $& \ion{N}{ii}  &$5007.3       $& \ion{O}{ii} &$4941.1-4943.0$& \ion{Si}{iii}&$3791.4-3806.8$\\
		\ion{H}{i} &$8392.2$& \ion{He}{i}  &$4471.5$& \ion{C}{ii}  &$6578.1-6582.9$& \ion{N}{ii}  &$5045.1       $& \ion{O}{iii}&$3703.4       $& \ion{Si}{iii}&$4552.6-4574.8$\\
		\ion{H}{i} &$8413.1$& \ion{He}{i}  &$4713.1$& \ion{C}{ii}  &$6783.9       $& \ion{N}{ii}  &$5666.6-5710.8$& \ion{O}{iii}&$3707.3-3715.1$& \ion{Si}{iii}&$4716.7       $\\
		\ion{H}{i} &$8437.8$& \ion{He}{i}  &$4921.9$& \ion{C}{iii} &$4056.1       $& \ion{N}{ii}  &$5931.9-5941.7$& \ion{O}{iii}&$3754.7-3791.3$& \ion{Si}{iii}&$4813.3-4829.0$\\
		\ion{H}{i} &$8467.0$& \ion{He}{i}  &$5015.7$& \ion{C}{iii} &$4068.9-4070.3$& \ion{N}{ii}  &$6482.1       $& \ion{O}{iii}&$3961.6       $& \ion{Si}{iii}&$5739.7       $\\
		\ion{H}{i} &$8502.3$& \ion{He}{i}  &$5047.7$& \ion{C}{iii} &$4152.5-4162.9$& \ion{N}{ii}  &$6610.6       $& \ion{O}{iii}&$4072.6-4089.3$& \ion{Si}{iii}&$7461.9-7466.3$\\
		\ion{H}{i} &$8545.2$& \ion{He}{i}  &$5875.6$& \ion{C}{iii} &$4186.9       $& \ion{N}{iii} &$3934.5-3938.5$& \ion{O}{iii}&$4366.5-4375.9$& \ion{Si}{iii}&$8262.6-8271.9$\\
		\ion{H}{i} &$8598.2$& \ion{He}{i}  &$6678.2$& \ion{C}{iii} &$4647.4-4651.5$& \ion{N}{iii} &$3998.6-4003.6$& \ion{O}{iii}&$4799.8       $& \ion{Si}{iii}&$9799.9       $\\
		\ion{H}{i} &$8664.8$& \ion{He}{i}  &$7065.2$& \ion{C}{iii} &$4663.6-4665.9$& \ion{N}{iii} &$4097.4-4103.4$& \ion{O}{iii}&$5268.3       $& \ion{Si}{iv} &$4088.9-4116.1$\\
		\ion{H}{i} &$8750.3$& \ion{He}{i}  &$7281.4$& \ion{C}{iii} &$5249.1       $& \ion{N}{iii} &$4195.8-4200.1$& \ion{O}{iii}&$5508.2       $& \ion{Si}{iv} &$4212.4       $\\
		\ion{H}{i} &$8862.6$& \ion{He}{ii} &$3796.3$& \ion{C}{iii} &$5253.6-5272.5$& \ion{N}{iii} &$4332.9-4345.7$& \ion{O}{iii}&$5592.3       $& \ion{Si}{iv} &$4950.1       $\\
		\ion{H}{i} &$9014.7$& \ion{He}{ii} &$3813.5$& \ion{C}{iii} &$5695.9       $& \ion{N}{iii} &$4379.1       $& \ion{O}{iii}&$7711.0       $& \ion{Si}{iv} &$6667.6-6701.3$\\
		\ion{H}{i} &$9228.8$& \ion{He}{ii} &$3833.8$& \ion{C}{iii} &$5826.4       $& \ion{N}{iii} &$4510.9-4547.3$& \ion{O}{iv} &$3560.4-3563.3$& \ion{Si}{iv} &$7047.9-7068.4$\\
		\ion{H}{i} &$9545.7$& \ion{He}{ii} &$3858.1$& \ion{C}{iii} &$6731.0-6744.3$& \ion{N}{iii} &$4527.9-4546.3$& \ion{O}{iv} &$3729.0-3736.9$& \ion{Si}{iv} &$8957.3       $\\
		           &        & \ion{He}{ii} &$3887.5$& \ion{C}{iii} &$7707.4       $& \ion{N}{iii} &$4634.1-4641.9$& \ion{O}{iv} &$3995.1       $& \ion{Si}{iv} &$9018.1       $\\
		           &        & \ion{He}{ii} &$3923.5$& \ion{C}{iii} &$8500.3       $& \ion{N}{iii} &$4858.7-4867.2$& \ion{O}{iv} &$4654.1       $&              & \\
		           &        & \ion{He}{ii} &$3968.4$& \ion{C}{iii} &$9701.1-9715.1$& \ion{N}{iii} &$5320.8-5352.5$& \ion{O}{iv} &$4813.2       $&              & \\
		           &        & \ion{He}{ii} &$4025.6$& \ion{C}{iv}  &$4646.6-4647.0$& \ion{N}{iii} &$6445.3-6487.8$& \ion{O}{iv} &$7004.1       $&              & \\
		           &        & \ion{He}{ii} &$4100.1$& \ion{C}{iv}  &$5016.6-5018.4$& \ion{N}{iii} &$9402.5-9424.5$& \ion{O}{iv} &$7032.3-7053.6$&              & \\
		           &        & \ion{He}{ii} &$4199.8$& \ion{C}{iv}  &$5801.3-5812.0$& \ion{N}{iv}  &$3747.5       $& \ion{O}{iv} &$9453.9-0402.4$&              & \\
		           &        & \ion{He}{ii} &$4338.7$& \ion{C}{iv}  &$6591.5-6592.6$& \ion{N}{iv}  &$4057.8       $& \ion{O}{v}  &$5114.1       $&              & \\
		           &        & \ion{He}{ii} &$4541.6$&              &               & \ion{N}{iv}  &$5200.4-5205.1$& \ion{O}{v}  &$6500.2       $&              & \\
		           &        & \ion{He}{ii} &$4685.7$&              &               & \ion{N}{iv}  &$5736.9       $&             &               &              & \\
		           &        & \ion{He}{ii} &$4859.3$&              &               & \ion{N}{iv}  &$5776.3-5784.8$&             &               &              & \\
		           &        & \ion{He}{ii} &$5411.5$&              &               & \ion{N}{iv}  &$6212.4-6219.9$&             &               &              & \\
		           &        & \ion{He}{ii} &$6074.2$&              &               & \ion{N}{iv}  &$7103.2-7129.2$&             &               &              & \\
		           &        & \ion{He}{ii} &$6118.3$&              &               & \ion{N}{iv}  &$7425.3       $&             &               &              & \\
		           &        & \ion{He}{ii} &$6170.7$&              &               & \ion{N}{iv}  &$9182.2-9223.0$&             &               &              & \\
		           &        & \ion{He}{ii} &$6233.8$&              &               & \ion{N}{v}   &$4603.7-4620.0$&             &               &              & \\
		           &        & \ion{He}{ii} &$6310.9$&              &               &              &               &             &               &              & \\
		           &        & \ion{He}{ii} &$6406.4$&              &               &              &               &             &               &              & \\
		           &        & \ion{He}{ii} &$6527.1$&              &               &              &               &             &               &              & \\
		           &        & \ion{He}{ii} &$6560.1$&              &               &              &               &             &               &              & \\
		           &        & \ion{He}{ii} &$6683.2$&              &               &              &               &             &               &              & \\
		           &        & \ion{He}{ii} &$6890.9$&              &               &              &               &             &               &              & \\
		           &        & \ion{He}{ii} &$7177.5$&              &               &              &               &             &               &              & \\
		           &        & \ion{He}{ii} &$7592.8$&              &               &              &               &             &               &              & \\
		           &        & \ion{He}{ii} &$8236.8$&              &               &              &               &             &               &              & \\
		           &        & \ion{He}{ii} &$9011.2$&              &               &              &               &             &               &              & \\
		           &        & \ion{He}{ii} &$9108.6$&              &               &              &               &             &               &              & \\
		           &        & \ion{He}{ii} &$9225.3$&              &               &              &               &             &               &              & \\
		           &        & \ion{He}{ii} &$9344.9$&              &               &              &               &             &               &              & \\
		           &        & \ion{He}{ii} &$9367.1$&              &               &              &               &             &               &              & \\
		           &        & \ion{He}{ii} &$9542.1$&              &               &              &               &             &               &              & \\
		           &        &              &        &              &               &              &               &             &               &              & \\
		\hline                               
	\end{tabular}                                
\end{table*}

\begin{figure}                                       
\begin{center}                                       
 \includegraphics[width=\columnwidth]{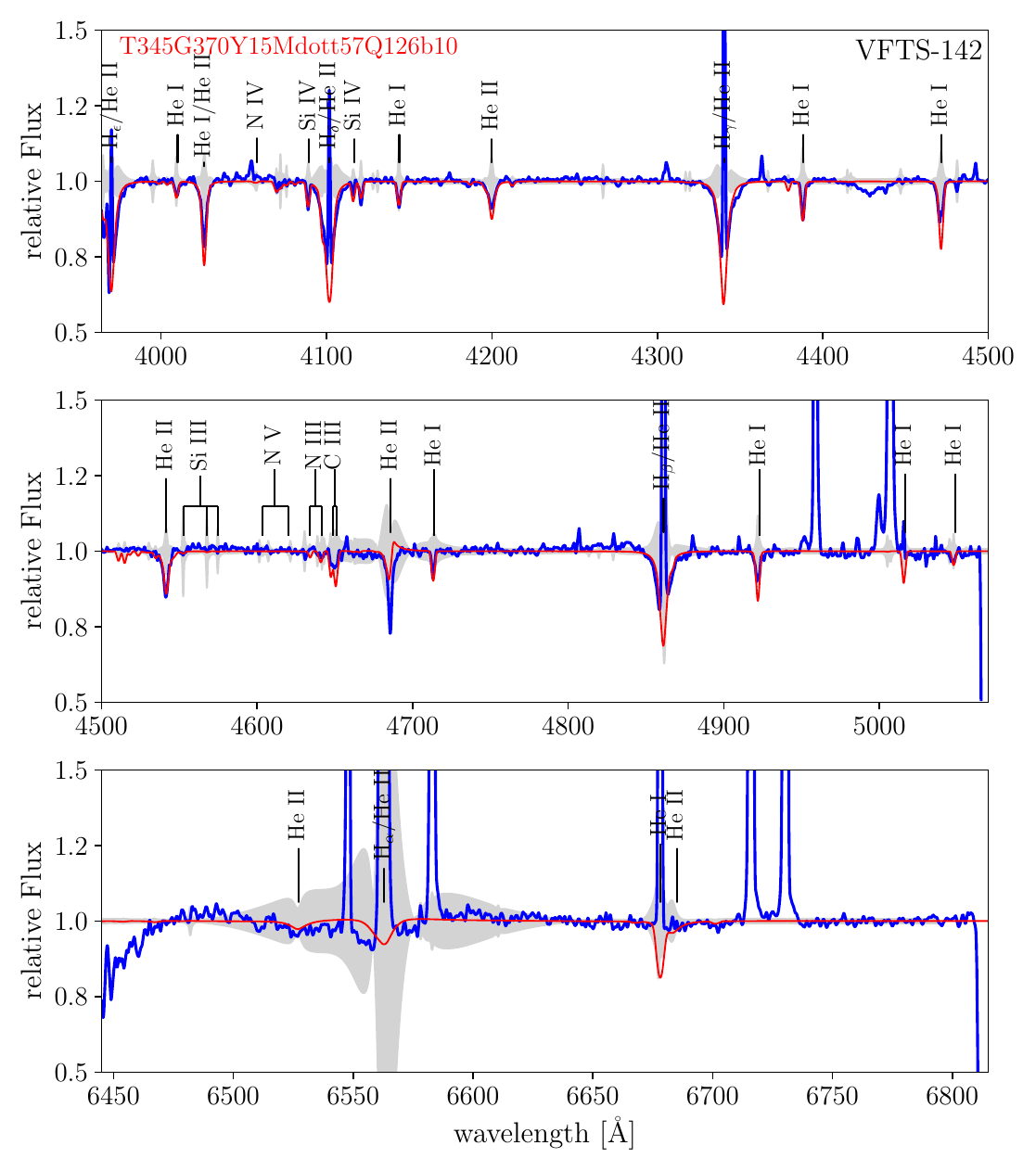} 
 \caption{VFTS~142: moderate nebular contamination. Blue solid line is the observation, red solid line the synthetic spectrum
and the grey shaded area is the square-root of the diagonal elements of the covariant-matrix calculated by the pipeline. Effective temperature is well reproduce while the surface gravity is 0.2 to 0.3~dex too low.}
   \label{af:142}
\end{center}
\end{figure}

\begin{figure}
\begin{center}
 \includegraphics[width=\columnwidth]{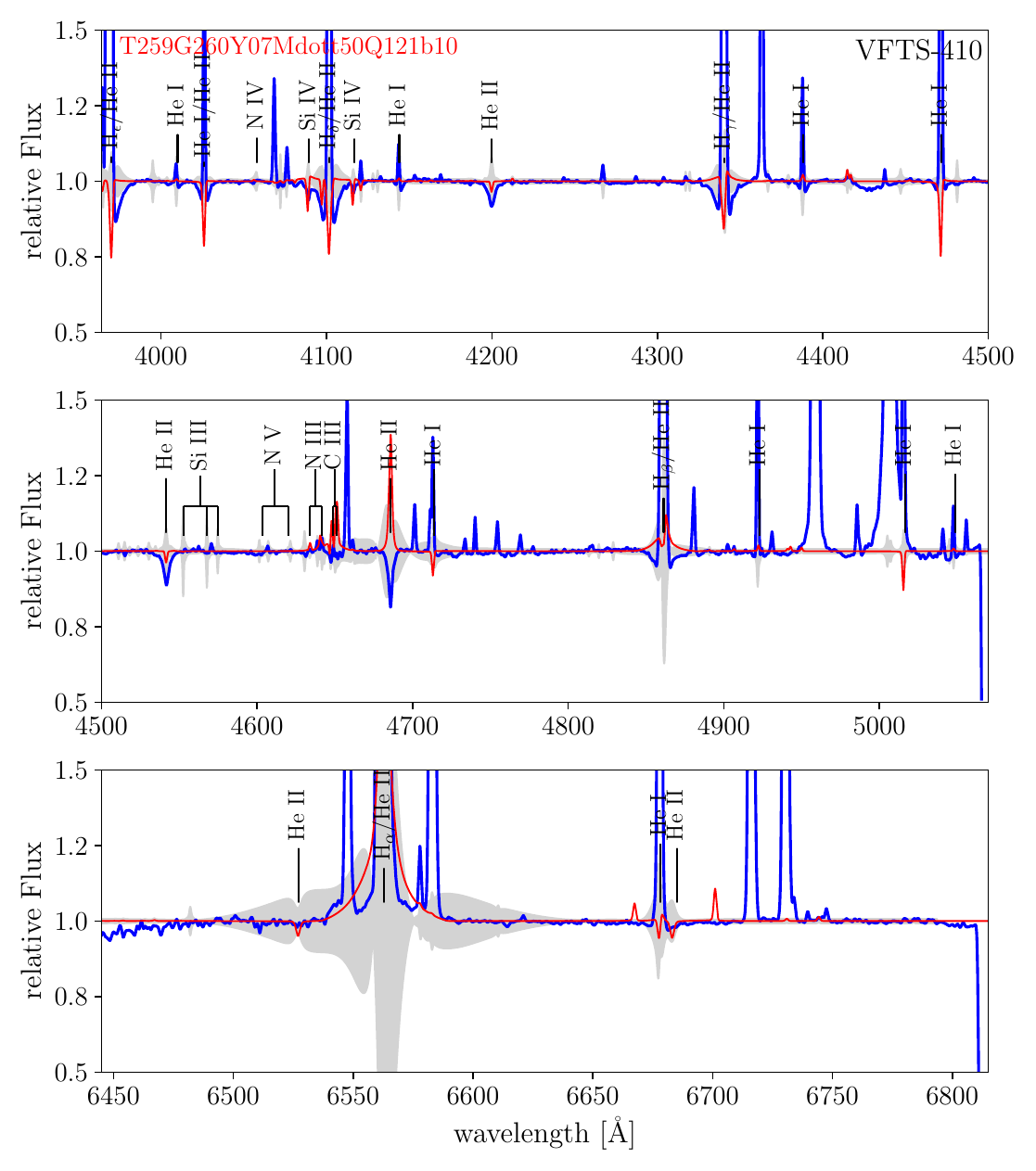} 
 \caption{VFTS~410: strong nebular contamination. Blue solid line is the observation, red solid line the synthetic spectrum
and the grey shaded area is the square-root of the diagonal elements of the covariant-matrix calculated by the pipeline. The pipeline is unable to reproduce the stellar parameters.}
   \label{af:410}
\end{center}
\end{figure}

\begin{figure}
\begin{center}
 \includegraphics[width=\columnwidth]{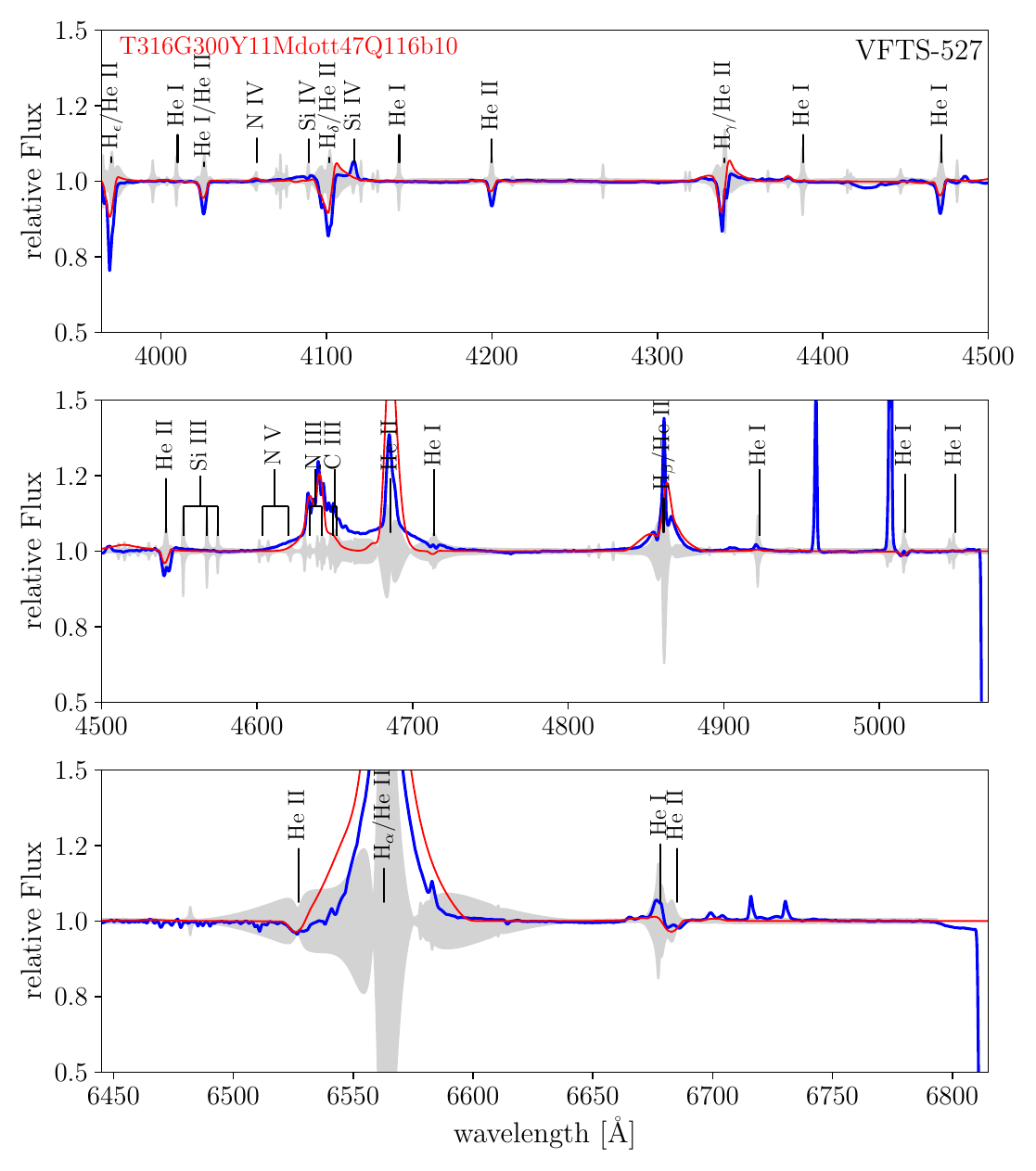} 
 \caption{VFTS~527: double-lined spectroscopic binary VFTS-527 \citep{taylor2011}. Blue solid line is the observation, red solid line the synthetic spectrum
and the grey shaded area is the square-root of the diagonal elements of the covariant-matrix calculated by the pipeline. Effective temperature and surface gravity of the primary are reproduced, but mass-loss rate is too high while helium abundance is too low due to the contribution of the colliding wind region largely effecting $\mathrm{H}_{\alpha}$ and $\mathrm{H}_{\beta}$.}
   \label{af:527}
\end{center}
\end{figure}

\onecolumn
\begin{landscape}
\setlength{\tabcolsep}{0.4pt}
\begin{table*}
\caption{Error-model uncertainties-matrix based on the VFTS analysis (\S~\ref{s:vfts}). The rank of the matrix has been reduced by merging elements together for visualisation purposes. The Balmer lines show the strongest correlations have been highlighted.}\label{at:err_mat}
\begin{tabular}{lcccc|r@{~}l|ccccc|r@{~}l|cccccccccccc|r@{~}l|cccccc|r@{~}l|cc}
\hline
&   &&&&\multicolumn{2}{|c|}{H$_{\delta}$} &      &     &&&&\multicolumn{2}{|c|}{H$_{\gamma}$}&      &      &      &      &      &      &      &      &      &      &&&\multicolumn{2}{|c|}{H$_{\beta}$}&      &      &     &&&&\multicolumn{2}{|c|}{H$_{\alpha}$}&      &\\
\hline
$\lambda$ [\AA]~&3987 & 4016 & 4045 & 4072 & ~~4089&4115~~& 4136 & 4160 & 4192 & 4233 & 4310 & ~~4325 & 4346~~ & 4378 & 4424 & 4474 & 4511 & 4532 & 4559 & 4613 & 4638 & 4651 & 4669 & 4717 & 4792 & ~~4824 & 4875~~ & 4914 & 4987 & 5008 & 5025 & 6447 & 6472 & ~~6520 & 6582~~ & 6643 & 6728 \\
3987 & \textbf{0.03} & 0.02 & 0.01 & -0.0 & 0.01 & 0.02 & 0.02 & 0.0 & -0.01 & 0.0 & 0.01 & 0.0 & 0.01 & 0.01 & 0.0 & 0.03 & 0.0 & -0.01 & 0.02 & 0.02 & 0.05 & -0.07 & -0.02 & 0.01 & 0.0 & 0.02 & 0.0 & 0.01 & 0.0 & 0.03 & 0.0 & 0.01 & 0.01 & 0.06 & -0.13 & 0.03 & -0.0 \\
4016 & 0.02 & \textbf{0.02} & 0.01 & -0.0 & 0.0 & 0.01 & 0.01 & 0.01 & 0.01 & 0.01 & 0.01 & -0.0 & 0.01 & 0.02 & 0.01 & 0.02 & 0.01 & 0.01 & 0.01 & 0.02 & 0.02 & -0.01 & 0.01 & 0.01 & 0.01 & -0.01 & 0.04 & 0.01 & 0.01 & 0.03 & 0.02 & 0.02 & 0.02 & -0.04 & 0.17 & 0.0 & 0.01 \\
4045 & 0.01 & 0.01 & \textbf{0.01} & -0.01 & 0.0 & 0.0 & 0.01 & 0.01 & 0.01 & 0.01 & 0.0 & -0.0 & 0.0 & 0.01 & 0.01 & 0.01 & 0.01 & 0.01 & 0.0 & 0.01 & 0.01 & 0.01 & 0.01 & 0.01 & 0.01 & -0.01 & 0.02 & 0.01 & 0.01 & 0.02 & 0.01 & 0.01 & 0.01 & -0.03 & 0.13 & -0.0 & 0.01 \\
\smallskip
4072 & -0.0 & -0.0 & -0.01 & \textbf{0.03} & -0.03 & -0.01 & 0.0 & -0.0 & -0.01 & -0.0 & 0.0 & -0.0 & -0.01 & -0.0 & 0.01 & 0.0 & -0.01 & -0.02 & 0.01 & 0.0 & 0.0 & -0.03 & -0.0 & -0.01 & -0.0 & -0.0 & 0.02 & 0.0 & 0.0 & 0.0 & 0.0 & 0.0 & -0.0 & -0.01 & 0.1 & -0.01 & -0.0 \\
4089 & 0.01 & 0.0 & 0.0 & -0.03 & \textbf{0.03} & 0.02 & -0.0 & -0.0 & 0.0 & -0.0 & -0.01 & 0.0 & 0.01 & 0.0 & -0.01 & 0.0 & 0.01 & 0.01 & -0.01 & -0.01 & 0.0 & 0.01 & -0.01 & 0.0 & -0.0 & 0.01 & -0.04 & -0.0 & -0.01 & -0.01 & -0.01 & -0.01 & 0.0 & 0.05 & -0.25 & 0.03 & -0.01 \\
\smallskip
4115 & 0.02 & 0.01 & 0.0 & -0.01 & 0.02 & \textbf{0.02} & 0.01 & -0.01 & -0.01 & -0.01 & -0.0 & 0.01 & 0.01 & 0.0 & -0.01 & 0.01 & -0.0 & -0.01 & 0.01 & -0.0 & 0.03 & -0.06 & -0.03 & 0.0 & -0.01 & 0.03 & -0.04 & 0.0 & -0.01 & -0.0 & -0.01 & -0.01 & 0.0 & 0.1 & -0.34 & 0.04 & -0.01 \\
4136 & 0.02 & 0.01 & 0.01 & 0.0 & -0.0 & 0.01 & \textbf{0.02} & 0.0 & -0.01 & 0.0 & 0.01 & 0.0 & 0.01 & 0.01 & 0.01 & 0.02 & 0.0 & -0.0 & 0.02 & 0.01 & 0.03 & -0.05 & -0.01 & 0.01 & 0.0 & 0.01 & 0.01 & 0.01 & 0.0 & 0.02 & 0.01 & 0.01 & 0.01 & 0.02 & -0.01 & 0.02 & 0.0 \\
4160 & 0.0 & 0.01 & 0.01 & -0.0 & -0.0 & -0.01 & 0.0 & \textbf{0.02} & 0.02 & 0.02 & 0.01 & -0.01 & -0.0 & 0.01 & 0.02 & 0.0 & 0.02 & 0.02 & 0.0 & 0.01 & -0.01 & 0.04 & 0.03 & 0.01 & 0.01 & -0.03 & 0.05 & 0.01 & 0.02 & 0.03 & 0.02 & 0.02 & 0.01 & -0.1 & 0.35 & -0.02 & 0.02 \\
4192 & -0.01 & 0.01 & 0.01 & -0.01 & 0.0 & -0.01 & -0.01 & 0.02 & \textbf{0.02} & 0.02 & 0.0 & -0.01 & -0.0 & 0.01 & 0.01 & -0.01 & 0.02 & 0.03 & -0.01 & 0.01 & -0.03 & 0.08 & 0.04 & 0.01 & 0.01 & -0.03 & 0.05 & 0.0 & 0.02 & 0.01 & 0.02 & 0.01 & 0.01 & -0.11 & 0.36 & -0.03 & 0.02 \\
4233 & 0.0 & 0.01 & 0.01 & -0.0 & -0.0 & -0.01 & 0.0 & 0.02 & 0.02 & \textbf{0.02} & 0.01 & -0.01 & -0.0 & 0.01 & 0.02 & 0.0 & 0.02 & 0.02 & 0.0 & 0.01 & -0.01 & 0.04 & 0.03 & 0.01 & 0.02 & -0.03 & 0.06 & 0.01 & 0.02 & 0.03 & 0.02 & 0.02 & 0.01 & -0.1 & 0.37 & -0.02 & 0.02 \\
\smallskip
4310 & 0.01 & 0.01 & 0.0 & 0.0 & -0.01 & -0.0 & 0.01 & 0.01 & 0.0 & 0.01 & \textbf{0.0} & -0.0 & -0.0 & 0.01 & 0.01 & 0.01 & 0.0 & 0.0 & 0.01 & 0.01 & 0.01 & -0.0 & 0.01 & 0.01 & 0.01 & -0.01 & 0.03 & 0.0 & 0.01 & 0.02 & 0.01 & 0.01 & 0.01 & -0.03 & 0.14 & -0.01 & 0.01 \\
4325 & 0.0 & -0.0 & -0.0 & -0.0 & 0.0 & 0.01 & 0.0 & -0.01 & -0.01 & -0.01 & -0.0 & \textbf{0.01} & 0.0 & -0.0 & -0.01 & 0.0 & -0.01 & -0.01 & 0.0 & -0.01 & 0.01 & -0.04 & -0.02 & -0.01 & -0.01 & 0.02 & -0.03 & -0.0 & -0.01 & -0.01 & -0.01 & -0.01 & -0.01 & 0.07 & -0.24 & 0.02 & -0.01 \\
\smallskip
4346 & 0.01 & 0.01 & 0.0 & -0.01 & 0.01 & 0.01 & 0.01 & -0.0 & -0.0 & -0.0 & -0.0 & 0.0 & \textbf{0.01} & 0.0 & -0.0 & 0.01 & -0.0 & -0.0 & 0.0 & 0.0 & 0.01 & -0.02 & -0.01 & 0.0 & -0.0 & 0.01 & -0.01 & 0.0 & -0.0 & 0.0 & -0.0 & -0.0 & 0.0 & 0.04 & -0.13 & 0.02 & -0.0 \\
4378 & 0.01 & 0.02 & 0.01 & -0.0 & 0.0 & 0.0 & 0.01 & 0.01 & 0.01 & 0.01 & 0.01 & -0.0 & 0.0 & \textbf{0.01} & 0.01 & 0.01 & 0.01 & 0.01 & 0.01 & 0.02 & 0.02 & -0.01 & 0.01 & 0.01 & 0.01 & -0.01 & 0.04 & 0.01 & 0.01 & 0.03 & 0.02 & 0.02 & 0.01 & -0.04 & 0.17 & -0.0 & 0.01 \\
4424 & 0.0 & 0.01 & 0.01 & 0.01 & -0.01 & -0.01 & 0.01 & 0.02 & 0.01 & 0.02 & 0.01 & -0.01 & -0.0 & 0.01 & \textbf{0.02} & 0.01 & 0.02 & 0.02 & 0.01 & 0.02 & -0.0 & 0.03 & 0.03 & 0.01 & 0.02 & -0.03 & 0.06 & 0.01 & 0.02 & 0.03 & 0.03 & 0.02 & 0.01 & -0.1 & 0.4 & -0.03 & 0.02 \\
4474 & 0.03 & 0.02 & 0.01 & 0.0 & 0.0 & 0.01 & 0.02 & 0.0 & -0.01 & 0.0 & 0.01 & 0.0 & 0.01 & 0.01 & 0.01 & \textbf{0.02} & 0.0 & -0.01 & 0.02 & 0.02 & 0.04 & -0.06 & -0.02 & 0.01 & 0.0 & 0.01 & 0.01 & 0.01 & 0.0 & 0.03 & 0.01 & 0.01 & 0.01 & 0.03 & -0.03 & 0.02 & 0.0 \\
4511 & 0.0 & 0.01 & 0.01 & -0.01 & 0.01 & -0.0 & 0.0 & 0.02 & 0.02 & 0.02 & 0.0 & -0.01 & -0.0 & 0.01 & 0.02 & 0.0 & \textbf{0.02} & 0.03 & -0.0 & 0.01 & -0.01 & 0.05 & 0.03 & 0.01 & 0.02 & -0.03 & 0.05 & 0.01 & 0.02 & 0.03 & 0.02 & 0.02 & 0.01 & -0.09 & 0.31 & -0.02 & 0.02 \\
4532 & -0.01 & 0.01 & 0.01 & -0.02 & 0.01 & -0.01 & -0.0 & 0.02 & 0.03 & 0.02 & 0.0 & -0.01 & -0.0 & 0.01 & 0.02 & -0.01 & 0.03 & \textbf{0.04} & -0.01 & 0.01 & -0.03 & 0.09 & 0.05 & 0.01 & 0.02 & -0.04 & 0.05 & 0.0 & 0.02 & 0.02 & 0.03 & 0.02 & 0.01 & -0.13 & 0.41 & -0.03 & 0.02 \\
4559 & 0.02 & 0.01 & 0.0 & 0.01 & -0.01 & 0.01 & 0.02 & 0.0 & -0.01 & 0.0 & 0.01 & 0.0 & 0.0 & 0.01 & 0.01 & 0.02 & -0.0 & -0.01 & \textbf{0.02} & 0.01 & 0.04 & -0.06 & -0.02 & 0.0 & 0.0 & 0.01 & 0.01 & 0.01 & 0.0 & 0.02 & 0.0 & 0.01 & 0.01 & 0.03 & -0.03 & 0.01 & -0.0 \\
4613 & 0.02 & 0.02 & 0.01 & 0.0 & -0.01 & -0.0 & 0.01 & 0.01 & 0.01 & 0.01 & 0.01 & -0.01 & 0.0 & 0.02 & 0.02 & 0.02 & 0.01 & 0.01 & 0.01 & \textbf{0.02} & 0.02 & -0.01 & 0.01 & 0.01 & 0.01 & -0.01 & 0.05 & 0.01 & 0.01 & 0.04 & 0.02 & 0.02 & 0.02 & -0.05 & 0.24 & -0.01 & 0.01 \\
4638 & 0.05 & 0.02 & 0.01 & 0.0 & 0.0 & 0.03 & 0.03 & -0.01 & -0.03 & -0.01 & 0.01 & 0.01 & 0.01 & 0.02 & -0.0 & 0.04 & -0.01 & -0.03 & 0.04 & 0.02 & \textbf{0.09} & -0.14 & -0.06 & 0.01 & -0.0 & 0.05 & -0.02 & 0.01 & -0.01 & 0.03 & -0.01 & 0.01 & 0.01 & 0.15 & -0.37 & 0.06 & -0.01 \\
4651 & -0.07 & -0.01 & 0.01 & -0.03 & 0.01 & -0.06 & -0.05 & 0.04 & 0.08 & 0.04 & -0.0 & -0.04 & -0.02 & -0.01 & 0.03 & -0.06 & 0.05 & 0.09 & -0.06 & -0.01 & -0.14 & \textbf{0.32} & 0.14 & 0.01 & 0.03 & -0.12 & 0.11 & -0.01 & 0.04 & -0.0 & 0.05 & 0.02 & 0.0 & -0.38 & 1.1 & -0.12 & 0.06 \\
4669 & -0.02 & 0.01 & 0.01 & -0.0 & -0.01 & -0.03 & -0.01 & 0.03 & 0.04 & 0.03 & 0.01 & -0.02 & -0.01 & 0.01 & 0.03 & -0.02 & 0.03 & 0.05 & -0.02 & 0.01 & -0.06 & 0.14 & \textbf{0.08} & 0.02 & 0.03 & -0.07 & 0.1 & 0.0 & 0.03 & 0.02 & 0.04 & 0.03 & 0.01 & -0.23 & 0.75 & -0.07 & 0.04 \\
4717 & 0.01 & 0.01 & 0.01 & -0.01 & 0.0 & 0.0 & 0.01 & 0.01 & 0.01 & 0.01 & 0.01 & -0.01 & 0.0 & 0.01 & 0.01 & 0.01 & 0.01 & 0.01 & 0.0 & 0.01 & 0.01 & 0.01 & 0.02 & \textbf{0.01} & 0.01 & -0.01 & 0.04 & 0.01 & 0.01 & 0.02 & 0.02 & 0.02 & 0.01 & -0.05 & 0.21 & -0.01 & 0.01 \\
\smallskip
4792 & 0.0 & 0.01 & 0.01 & -0.0 & -0.0 & -0.01 & 0.0 & 0.01 & 0.01 & 0.02 & 0.01 & -0.01 & -0.0 & 0.01 & 0.02 & 0.0 & 0.02 & 0.02 & 0.0 & 0.01 & -0.0 & 0.03 & 0.03 & 0.01 & \textbf{0.01} & -0.02 & 0.05 & 0.01 & 0.01 & 0.03 & 0.02 & 0.02 & 0.01 & -0.09 & 0.31 & -0.02 & 0.02 \\
4824 & 0.02 & -0.01 & -0.01 & -0.0 & 0.01 & 0.03 & 0.01 & -0.03 & -0.03 & -0.03 & -0.01 & 0.02 & 0.01 & -0.01 & -0.03 & 0.01 & -0.03 & -0.04 & 0.01 & -0.01 & 0.05 & -0.12 & -0.07 & -0.01 & -0.02 & \textbf{0.06} & -0.09 & -0.0 & -0.03 & -0.02 & -0.04 & -0.02 & -0.01 & 0.21 & -0.69 & 0.06 & -0.03 \\
\smallskip
4875 & 0.0 & 0.04 & 0.02 & 0.02 & -0.04 & -0.04 & 0.01 & 0.05 & 0.05 & 0.06 & 0.03 & -0.03 & -0.01 & 0.04 & 0.06 & 0.01 & 0.05 & 0.05 & 0.01 & 0.05 & -0.02 & 0.11 & 0.1 & 0.04 & 0.05 & -0.09 & \textbf{0.2} & 0.02 & 0.05 & 0.09 & 0.08 & 0.06 & 0.04 & -0.33 & 1.25 & -0.08 & 0.06 \\
4914 & 0.01 & 0.01 & 0.01 & 0.0 & -0.0 & 0.0 & 0.01 & 0.01 & 0.0 & 0.01 & 0.0 & -0.0 & 0.0 & 0.01 & 0.01 & 0.01 & 0.01 & 0.0 & 0.01 & 0.01 & 0.01 & -0.01 & 0.0 & 0.01 & 0.01 & -0.0 & 0.02 & \textbf{0.01} & 0.01 & 0.02 & 0.01 & 0.01 & 0.01 & -0.02 & 0.11 & 0.0 & 0.01 \\
4987 & 0.0 & 0.01 & 0.01 & 0.0 & -0.01 & -0.01 & 0.0 & 0.02 & 0.02 & 0.02 & 0.01 & -0.01 & -0.0 & 0.01 & 0.02 & 0.0 & 0.02 & 0.02 & 0.0 & 0.01 & -0.01 & 0.04 & 0.03 & 0.01 & 0.01 & -0.03 & 0.05 & 0.01 & \textbf{0.02} & 0.03 & 0.02 & 0.02 & 0.01 & -0.09 & 0.35 & -0.02 & 0.02 \\
5008 & 0.03 & 0.03 & 0.02 & 0.0 & -0.01 & -0.0 & 0.02 & 0.03 & 0.01 & 0.03 & 0.02 & -0.01 & 0.0 & 0.03 & 0.03 & 0.03 & 0.03 & 0.02 & 0.02 & 0.04 & 0.03 & -0.0 & 0.02 & 0.02 & 0.03 & -0.02 & 0.09 & 0.02 & 0.03 & \textbf{0.06} & 0.04 & 0.04 & 0.03 & -0.1 & 0.46 & -0.02 & 0.03 \\
5025 & 0.0 & 0.02 & 0.01 & 0.0 & -0.01 & -0.01 & 0.01 & 0.02 & 0.02 & 0.02 & 0.01 & -0.01 & -0.0 & 0.02 & 0.03 & 0.01 & 0.02 & 0.03 & 0.0 & 0.02 & -0.01 & 0.05 & 0.04 & 0.02 & 0.02 & -0.04 & 0.08 & 0.01 & 0.02 & 0.04 & \textbf{0.03} & 0.03 & 0.02 & -0.14 & 0.51 & -0.03 & 0.03 \\
6447 & 0.01 & 0.02 & 0.01 & 0.0 & -0.01 & -0.01 & 0.01 & 0.02 & 0.01 & 0.02 & 0.01 & -0.01 & -0.0 & 0.02 & 0.02 & 0.01 & 0.02 & 0.02 & 0.01 & 0.02 & 0.01 & 0.02 & 0.03 & 0.02 & 0.02 & -0.02 & 0.06 & 0.01 & 0.02 & 0.04 & 0.03 & \textbf{0.02} & 0.02 & -0.09 & 0.36 & -0.02 & 0.02 \\
\smallskip
6472 & 0.01 & 0.02 & 0.01 & -0.0 & 0.0 & 0.0 & 0.01 & 0.01 & 0.01 & 0.01 & 0.01 & -0.01 & 0.0 & 0.01 & 0.01 & 0.01 & 0.01 & 0.01 & 0.01 & 0.02 & 0.01 & 0.0 & 0.01 & 0.01 & 0.01 & -0.01 & 0.04 & 0.01 & 0.01 & 0.03 & 0.02 & 0.02 & \textbf{0.01} & -0.05 & 0.2 & -0.0 & 0.01 \\
6520 & 0.06 & -0.04 & -0.03 & -0.01 & 0.05 & 0.1 & 0.02 & -0.1 & -0.11 & -0.1 & -0.03 & 0.07 & 0.04 & -0.04 & -0.1 & 0.03 & -0.09 & -0.13 & 0.03 & -0.05 & 0.15 & -0.38 & -0.23 & -0.05 & -0.09 & 0.21 & -0.33 & -0.02 & -0.09 & -0.1 & -0.14 & -0.09 & -0.05 & \textbf{0.73} & -2.46 & 0.21 & -0.12 \\
\smallskip
6582 & -0.13 & 0.17 & 0.13 & 0.1 & -0.25 & -0.34 & -0.01 & 0.35 & 0.36 & 0.37 & 0.14 & -0.24 & -0.13 & 0.17 & 0.4 & -0.03 & 0.31 & 0.41 & -0.03 & 0.24 & -0.37 & 1.1 & 0.75 & 0.21 & 0.31 & -0.69 & 1.25 & 0.11 & 0.35 & 0.46 & 0.51 & 0.36 & 0.2 & -2.46 & \textbf{8.74} & -0.7 & 0.41 \\
6643 & 0.03 & 0.0 & -0.0 & -0.01 & 0.03 & 0.04 & 0.02 & -0.02 & -0.03 & -0.02 & -0.01 & 0.02 & 0.02 & -0.0 & -0.03 & 0.02 & -0.02 & -0.03 & 0.01 & -0.01 & 0.06 & -0.12 & -0.07 & -0.01 & -0.02 & 0.06 & -0.08 & 0.0 & -0.02 & -0.02 & -0.03 & -0.02 & -0.0 & 0.21 & -0.7 & \textbf{0.07} & -0.03 \\
6728 & -0.0 & 0.01 & 0.01 & -0.0 & -0.01 & -0.01 & 0.0 & 0.02 & 0.02 & 0.02 & 0.01 & -0.01 & -0.0 & 0.01 & 0.02 & 0.0 & 0.02 & 0.02 & -0.0 & 0.01 & -0.01 & 0.06 & 0.04 & 0.01 & 0.02 & -0.03 & 0.06 & 0.01 & 0.02 & 0.03 & 0.03 & 0.02 & 0.01 & -0.12 & 0.41 & -0.03 & \textbf{0.02} \\
\hline
\end{tabular}
\end{table*}
\end{landscape}
\onecolumn
\begin{spacing}{1.1}
\begin{center}
\begin{longtable}{lcccl}
	\caption{Stellar parameters of the VLT/MUSE sources \citep{castro2018}, which must be verified before using them for scientific purposes (e.g. plots in the supplementary online material). A very basic automated quality control has been performed, comments column: grid boundary, derived stellar parameters are located at the edge of the $T_{\rm eff}-\log g$ parameter space of the stellar atmosphere grid; bad fit, large miss match between \ion{He}{i} and {\sc ii} lines; no photometry, insufficient photometric data were available to derive bolometric luminosities.}
	\label{at:muse_sp}\\
		\hline
		\multicolumn{1}{l}{ID} &  
    	\multicolumn{1}{c}{$T_{\rm eff}$}  &  
    	\multicolumn{1}{c}{$\log g$}    &    
    	\multicolumn{1}{c}{$\log L$}   &   
    	\multicolumn{1}{l}{comments} \\
    	\multicolumn{1}{l}{\citep{castro2018}} &   
    	\multicolumn{1}{c}{[kK]}  &  
    	\multicolumn{1}{c}{[g/cm$^2$]}    &    
    	\multicolumn{1}{c}{[$L/L_{\odot}$]}   &   
    	\multicolumn{1}{l}{} \\
		\hline
		\endfirsthead
		\multicolumn{4}{c}
		{\tablename\ \thetable\ -- \textit{Continued}} \\
		\hline
		\multicolumn{1}{l}{ID} &  
    	\multicolumn{1}{c}{$T_{\rm eff}$}  &  
    	\multicolumn{1}{c}{$\log g$}    &    
    	\multicolumn{1}{c}{$\log L$}   &   
    	\multicolumn{1}{l}{comments} \\
    	\multicolumn{1}{l}{\citep{castro2018}} &   
    	\multicolumn{1}{c}{[kK]}  &  
    	\multicolumn{1}{c}{[g/cm$^2$]}    &    
    	\multicolumn{1}{c}{[$L/L_{\odot}$]}   &   
    	\multicolumn{1}{l}{} \\
		\hline
		\endhead
		\hline 
		178 & $21.22^{+0.96}_{-0.65}$ & $2.99^{+0.25}_{-0.21}$ & --                      & no photometry\\
		195 & $37.63^{+0.85}_{-0.81}$ & $3.84^{+0.12}_{-0.11}$ & $4.91^{+0.04}_{-0.04 }$ & \\
		225 & $25.9^{+1.5}_{   }$ & $2.62^{+0.07}_{-0.02}$ & $3.64^{+0.14}_{-0.12     }$ & grid boundary, bad fit\\
		249 & $26.64^{+0.54}_{-0.46}$ & $3.6^{+0.24}_{-0.25}$ & $4.01^{+0.59}_{-0.59  }$ & \\
		276 & $37.67^{+1.08}_{-0.81}$ & $3.71^{+0.1}_{-0.08}$ & $5.23^{+0.06}_{-0.05  }$ & \\
		290 & $36.75^{+0.77}_{-0.69}$ & $3.81^{+0.1}_{-0.11}$ & $5.34^{+0.04}_{-0.04  }$ & \\
		322 & $21.61^{+5.27}_{-0.73}$ & $3.81^{+0.25}_{-0.16}$ & $3.46^{+0.36}_{-0.27 }$ & \\
		367 & $22.95^{+1.31}_{-1.31}$ & $3.82^{+0.08}_{-0.22}$ & $3.84^{+0.14}_{-0.14 }$ & \\
		375 & $34.6^{+0.65}_{-1.42}$ & $3.4^{+0.08}_{-0.15}$ & $4.73^{+0.03}_{-0.07   }$ & \\
		410 & $32.52^{+0.77}_{-1.23}$ & $3.19^{+0.1}_{-0.14}$ & $4.2^{+0.05}_{-0.07   }$ & \\
		466 & $36.33^{+6.11}_{-6.23}$ & $3.09^{+0.32}_{-0.37}$ & $4.97^{+0.22}_{-0.23 }$ & bad fit\\
		489 & $21.41^{+0.81}_{-0.58}$ & $3.8^{+0.04}_{-0.06}$ & $3.41^{+0.22}_{-0.22  }$ & \\
		501 & $33.44^{+2.46}_{-1.42}$ & $3.41^{+0.33}_{-0.19}$ & $4.29^{+0.12}_{-0.08 }$ & \\
		520 & $20.53^{+0.46}_{-0.46}$ & $2.59^{+0.09}_{-0.09}$ & $3.57^{+0.13}_{-0.13 }$ & \\
		543 & $35.64^{+2.61}_{-1.88}$ & $3.7^{+0.41}_{-0.27}$ & $4.48^{+0.11}_{-0.08  }$ & \\
		549 & $23.14^{+3.34}_{-1.54}$ & $3.9^{+0.11}_{-0.09}$ & $4.19^{+0.2}_{-0.1    }$ & \\
		605 & $21.26^{+4.88}_{-0.5}$ & $3.8^{+0.23}_{-0.08}$ & $3.68^{+0.3}_{-0.13    }$ & \\
		689 & $43.55^{+2.31}_{-2.04}$ & $3.71^{+0.1}_{-0.07}$ & $4.92^{+0.08}_{-0.08  }$ & \\
		710 & $32.68^{+1.81}_{-1.19}$ & $3.21^{+0.15}_{-0.13}$ & $4.27^{+0.09}_{-0.07 }$ & \\
		739 & $20.8^{+1.23}_{-1.08}$ & $3.79^{+0.07}_{-0.18}$ & $3.74^{+0.12}_{-0.11  }$ & \\
		762 & $27.37^{+0.73}_{-0.69}$ & $2.72^{+0.05}_{-0.06}$ & $5.08^{+0.06}_{-0.06 }$ & \\
		774 & $39.83^{+0.65}_{-0.65}$ & $3.49^{+0.06}_{-0.06}$ & $5.49^{+0.03}_{-0.03 }$ & \\
		805 & $33.25^{+3.34}_{-1.69}$ & $3.48^{+0.45}_{-0.27}$ & $4.55^{+0.17}_{-0.13 }$ & \\
		813 & $40.82^{+0.73}_{-1.27}$ & $4.37^{+0.03}_{-0.67}$ & $4.95^{+0.04}_{-0.05 }$ & bad fit\\
		830 & $36.83^{+1.81}_{-1.23}$ & $3.88^{+0.22}_{-0.21}$ & $5.41^{+0.08}_{-0.06 }$ & \\
		835 & $18.72^{+0.73}_{-0.92}$ & $3.6^{+0.11}_{-0.18}$ & $3.43^{+0.2}_{-0.21   }$ & \\
		888 & $29.02^{+1.35}_{-2.34}$ & $4.08^{+0.05}_{-0.1}$ & $4.39^{+0.08}_{-0.12  }$ & \\
		898 & $26.95^{+4.11}_{-5.42}$ & $4.0^{+0.26}_{-0.21}$ & $3.78^{+0.29}_{-0.32  }$ & \\
		909 & $33.41^{+0.69}_{-1.04}$ & $3.5^{+0.09}_{-0.11}$ & $4.5^{+0.05}_{-0.06   }$ & \\
		911 & $33.33^{+1.27}_{-0.88}$ & $4.19^{+0.06}_{-0.16}$ & $4.9^{+0.1}_{-0.09   }$ & \\
		912 & $30.14^{+1.08}_{-1.04}$ & $4.09^{+0.05}_{-0.1}$ & $4.22^{+0.18}_{-0.18  }$ & \\
		920 & $17.8^{+0.5}_{   }$ & $3.69^{+0.08}_{-0.07}$ & $3.23^{+1.07}_{-1.07     }$ & \\
		935 & $28.91^{+1.23}_{-1.08}$ & $4.07^{+0.05}_{-0.09}$ & $4.83^{+0.41}_{-0.41 }$ & \\
		955 & $38.36^{+0.77}_{-0.85}$ & $4.28^{+0.06}_{-0.28}$ & $4.89^{+0.04}_{-0.05 }$ & \\
		960 & $25.26^{+4.65}_{-7.46}$ & $2.52^{+1.13}_{-0.52}$ & $3.76^{+0.35}_{-0.41 }$ & bad fit\\
		963 & $44.71^{+3.34}_{-2.54}$ & $3.82^{+0.15}_{-0.09}$ & $5.63^{+0.12}_{-0.09 }$ & \\
		1008& $38.67^{+0.69}_{-0.77}$ & $4.0^{+0.16}_{-0.14}$ & $5.08^{+0.05}_{-0.05  }$ & \\
		1040& $35.48^{+0.81}_{-0.58}$ & $3.93^{+0.25}_{-0.12}$ & $4.77^{+0.05}_{-0.04 }$ & \\
		1068& $44.78^{+3.0}_{-1.27}$ & $3.7^{+0.13}_{-0.09}$ & $5.48^{+0.11}_{-0.06   }$ & \\
		1138& $35.6^{+0.54}_{-0.62}$ & $4.13^{+0.11}_{-0.14}$ & $4.46^{+0.14}_{-0.14  }$ & \\
		1143& $22.03^{+0.96}_{-1.19}$ & $3.84^{+0.1}_{-0.14}$ & $4.27^{+0.41}_{-0.41  }$ & \\
		1152& $25.99^{+0.85}_{-1.81}$ & $3.7^{+0.2}_{-0.2}$ & $4.51^{+0.05}_{-0.11    }$ & \\
		1177& $35.44^{+1.04}_{-1.15}$ & $3.61^{+0.16}_{-0.14}$ & $5.32^{+0.05}_{-0.05 }$ & \\
		1184& $31.56^{+0.54}_{-0.46}$ & $3.3^{+0.06}_{-0.07}$ & $5.31^{+0.1}_{-0.1    }$ & \\
		1199& $56.2^{    }_{-1.61}$ & $4.01^{+0.05}_{-0.07}$ & $6.33^{+0.5}_{-0.5     }$ & grid boundary\\
		1210& $26.64^{+0.85}_{-0.38}$ & $4.0^{+0.04}_{-0.05}$ & $3.87^{+0.23}_{-0.22  }$ & \\
		1223& $34.48^{+0.54}_{-0.81}$ & $3.35^{+0.09}_{-0.09}$ & $5.45^{+0.03}_{-0.04 }$ & \\
		1226& $18.34^{+0.35}_{-0.54}$ & $3.62^{+0.09}_{-0.17}$ & $3.36^{+0.03}_{-0.05 }$ & \\
		1274& $35.64^{+1.08}_{-1.11}$ & $3.6^{+0.13}_{-0.11}$ & $5.62^{+0.05}_{-0.05  }$ & \\
		1275& $31.64^{+1.35}_{-1.04}$ & $3.2^{+0.27}_{-0.13}$ & $4.63^{+0.36}_{-0.36  }$ & \\
		1276& $25.3^{+0.73}_{-2.31}$ & $3.83^{+0.18}_{-0.32}$ & $4.44^{+0.67}_{-0.67  }$ & \\
		1279& $27.56^{+1.31}_{-0.85}$ & $2.72^{+0.06}_{-0.06}$ & $5.43^{+0.08}_{-0.06 }$ & \\
		1282& $23.45^{+6.65}_{-3.46}$ & $4.05^{+0.14}_{-0.13}$ & $3.83^{+0.74}_{-0.73 }$ & \\
		1293& $31.56^{+0.54}_{-0.46}$ & $3.3^{+0.06}_{-0.07}$ & $5.65^{+0.03}_{-0.03  }$ & bad fit\\
		1297& $26.64^{+0.85}_{-0.46}$ & $3.99^{+0.05}_{-0.06}$ & $4.36^{+0.05}_{-0.03 }$ & \\
		1324& $32.37^{+1.96}_{-1.54}$ & $4.17^{+0.06}_{-0.16}$ & $3.97^{+0.61}_{-0.61 }$ & \\
		1325& $38.83^{+2.31}_{-0.85}$ & $3.71^{+0.13}_{-0.09}$ & $5.07^{+0.69}_{-0.69 }$ & \\
		1334& $34.56^{+0.77}_{-0.62}$ & $3.6^{+0.1}_{-0.08}$ & $4.72^{+0.05}_{-0.05   }$ & \\
		1340& $19.41^{+0.42}_{-0.38}$ & $3.61^{+0.08}_{-0.08}$ & $3.96^{+1.13}_{-1.13 }$ & bad fit\\
		1346& $36.83^{+1.11}_{-0.88}$ & $3.99^{+0.17}_{-0.15}$ & $5.44^{+0.05}_{-0.04 }$ & \\
		1370& $27.95^{+2.46}_{-3.69}$ & $4.09^{+0.06}_{-0.08}$ & $4.13^{+0.18}_{-0.22 }$ & \\
		1373& $37.6^{+1.65}_{-1.23}$ & $3.69^{+0.15}_{-0.13}$ & $5.6^{+0.07}_{-0.05   }$ & \\
		1387& $18.45^{+0.35}_{-0.46}$ & $3.68^{+0.04}_{-0.17}$ & $3.98^{+0.9}_{-0.9   }$ & \\
		1390& $26.64^{+0.46}_{-3.38}$ & $3.9^{+0.1}_{-0.13}$ & $4.24^{+0.2}_{-0.25    }$ & \\
		1398& $32.56^{+1.54}_{-1.08}$ & $3.43^{+0.25}_{-0.15}$ & $4.96^{+0.09}_{-0.07 }$ & \\
		1399& $19.34^{+0.42}_{-0.38}$ & $3.29^{+0.17}_{-0.16}$ & $4.79^{+0.04}_{-0.03 }$ & \\
		1401& $40.75^{+1.42}_{-2.08}$ & $3.9^{+0.19}_{-0.14}$ & $5.42^{+0.66}_{-0.66  }$ & \\
		1405& $25.99^{+1.69}_{-0.85}$ & $2.62^{+0.11}_{-0.07}$ & $4.61^{+0.13}_{-0.1  }$ & \\
		1409& $35.6^{+1.42}_{-1.19}$ & $3.8^{+0.46}_{-0.22}$ & $5.31^{+0.18}_{-0.17   }$ & \\
		1415& $44.71^{+0.69}_{-2.92}$ & $3.59^{+0.06}_{-0.14}$ & $5.09^{+1.06}_{-1.06 }$ & \\
		1423& $44.71^{+0.81}_{-1.04}$ & $3.59^{+0.07}_{-0.07}$ & $5.87^{+0.29}_{-0.29 }$ & \\
		1428& $29.14^{+2.15}_{-1.42}$ & $4.08^{+0.07}_{-0.09}$ & $4.12^{+0.41}_{-0.4  }$ & \\
		1433& $44.86^{+1.61}_{-1.0}$ & $3.7^{+0.07}_{-0.07}$ & $5.86^{+0.06}_{-0.04   }$ & \\
		1434& $29.83^{+1.42}_{-1.5}$ & $4.1^{+0.05}_{-0.09}$ & $4.11^{+0.14}_{-0.14   }$ & \\
		1439& $36.44^{+1.0}_{-1.27}$ & $3.76^{+0.19}_{-0.2}$ & $5.11^{+0.05}_{-0.06   }$ & \\
		1445& $36.56^{+1.5}_{-1.35}$ & $4.28^{+0.06}_{-0.14}$ & $4.95^{+0.62}_{-0.62  }$ & \\
		1447& $43.44^{+1.69}_{-1.73}$ & $4.4^{+0.06}_{-0.14}$ & $5.66^{+0.08}_{-0.08  }$ & \\
		1453& $21.64^{+0.88}_{-1.0}$ & $3.81^{+0.04}_{-0.04}$ & $4.09^{+0.54}_{-0.54  }$ & \\
		1456& $21.14^{+1.08}_{-3.34}$ & $3.26^{+0.39}_{-0.27}$ & $3.86^{+0.56}_{-0.56 }$ & \\
		1457& $17.8^{+0.54}_{   }$ & $3.59^{+0.05}_{-0.05}$ & $3.67^{+0.05}_{-0.0     }$ & grid boundary\\
		1459& $37.6^{+0.69}_{-0.77}$ & $3.4^{+0.07}_{-0.08}$ & $5.86^{+0.28}_{-0.28   }$ & \\
		1464& $35.44^{+0.69}_{-0.73}$ & $4.07^{+0.16}_{-0.14}$ & $4.35^{+0.12}_{-0.12 }$ & \\
		1472& $37.94^{+1.38}_{-1.65}$ & $3.82^{+0.29}_{-0.18}$ & $5.14^{+0.13}_{-0.13 }$ & \\
		1494& $51.47^{+3.96}_{-13.68}$ & $3.9^{+0.15}_{-0.8}$ & $5.97^{+0.86}_{-0.86  }$ & \\
		1503& $36.48^{+1.38}_{-1.27}$ & $4.28^{+0.05}_{-0.13}$ & $5.16^{+0.71}_{-0.71 }$ & \\
		1504& $27.22^{+3.11}_{-1.69}$ & $3.98^{+0.12}_{-0.35}$ & $4.22^{+0.27}_{-0.25 }$ & \\
		1505& $26.22^{+1.19}_{-2.0}$ & $3.98^{+0.05}_{-0.05}$ & $4.38^{+1.05}_{-1.05  }$ & \\
		1522& $36.33^{+1.08}_{-1.38}$ & $4.27^{+0.05}_{-0.12}$ & $5.12^{+0.52}_{-0.52 }$ & \\
		1523& $36.67^{+1.5}_{-0.96}$ & $4.24^{+0.09}_{-0.19}$ & $5.29^{+0.47}_{-0.47  }$ & \\
		1527& $36.71^{+0.85}_{-0.65}$ & $4.0^{+0.17}_{-0.12}$ & $5.26^{+0.95}_{-0.95  }$ & \\
		1535& $35.48^{+0.62}_{-0.73}$ & $4.06^{+0.17}_{-0.17}$ & $5.7^{+0.09}_{-0.09  }$ & \\
		1537& $35.87^{+1.42}_{-1.0}$ & $3.77^{+0.23}_{-0.15}$ & $5.33^{+0.35}_{-0.35  }$ & \\
		1572& $40.98^{+1.0}_{-1.88}$ & $4.38^{+0.04}_{-0.2}$ & $5.64^{+0.04}_{-0.08   }$ & \\
		1580& $46.01^{+1.19}_{-3.31}$ & $3.81^{+0.08}_{-0.07}$ & $5.67^{+0.64}_{-0.64 }$ & \\
		1607& $31.52^{+1.19}_{-0.85}$ & $3.5^{+0.17}_{-0.17}$ & $4.67^{+0.2}_{-0.19   }$ & \\
		1625& $35.52^{+0.88}_{-6.84}$ & $3.72^{+0.36}_{-0.16}$ & $5.19^{+0.26}_{-0.33 }$ & \\
		1632& $35.44^{+0.65}_{-0.96}$ & $3.69^{+0.1}_{-0.13}$ & $6.36^{+1.14}_{-1.14  }$ & \\
		1641& $22.18^{+1.27}_{-1.11}$ & $3.51^{+0.28}_{-0.25}$ & $4.41^{+0.09}_{-0.08 }$ & \\
		1651& $25.1^{+0.65}_{   }$ & $2.62^{+0.03}_{-0.02}$ & $4.24^{+0.08}_{-0.07    }$ & grid boundary, bad fit\\
		1682& $29.02^{+0.5}_{-3.88}$ & $4.0^{+0.07}_{-0.12}$ & $3.96^{+0.7}_{-0.7     }$ & \\
		1689& $30.72^{+0.65}_{-0.81}$ & $3.1^{+0.08}_{-0.1}$ & $5.78^{+0.04}_{-0.05   }$ & \\
		1699& $39.79^{+0.96}_{-1.04}$ & $3.72^{+0.19}_{-0.14}$ & $5.22^{+0.05}_{-0.05 }$ & \\
		1700& $36.06^{+1.19}_{-0.92}$ & $4.17^{+0.14}_{-0.29}$ & $5.52^{+0.05}_{-0.04 }$ & \\
		1714& $42.25^{+1.73}_{-2.42}$ & $4.21^{+0.2}_{-0.26}$ & $5.39^{+0.14}_{-0.15  }$ & \\
		1725& $27.4^{+0.41}_{-1.5}$ & $2.8^{+0.06}_{-0.13}$ & $4.82^{+0.99}_{-0.99    }$ & bad fit\\
		1749& $51.82^{+2.04}_{-1.77}$ & $4.01^{+0.1}_{-0.08}$ & $6.16^{+0.07}_{-0.06  }$ & \\
		1763& $36.63^{+1.58}_{-1.31}$ & $3.72^{+0.17}_{-0.14}$ & $5.6^{+0.59}_{-0.59  }$ & \\
		1764& $23.03^{+1.88}_{-1.31}$ & $3.79^{+0.11}_{-0.26}$ & $3.88^{+0.14}_{-0.11 }$ & \\
		1768& $26.64^{+1.23}_{-3.84}$ & $4.0^{+0.06}_{-0.1}$ & $4.3^{+0.18}_{-0.25    }$ & \\
		1778& $30.87^{+1.42}_{-1.23}$ & $4.11^{+0.07}_{-0.1}$ & $5.49^{+0.16}_{-0.16  }$ & \\
		1826& $37.63^{+1.15}_{-0.77}$ & $4.29^{+0.06}_{-0.24}$ & $4.38^{+0.21}_{-0.2  }$ & \\
		1827& $39.79^{+0.85}_{-0.85}$ & $3.73^{+0.16}_{-0.09}$ & $5.62^{+0.04}_{-0.04 }$ & \\
		1847& $18.53^{+0.58}_{-0.46}$ & $3.68^{+0.04}_{-0.15}$ & $2.82^{+1.24}_{-1.24 }$ & \\
		1857& $37.37^{+0.77}_{-0.88}$ & $4.22^{+0.1}_{-0.11}$ & $6.23^{+0.22}_{-0.23  }$ & \\
		1858& $25.95^{+1.38}_{-1.81}$ & $3.96^{+0.07}_{-0.07}$ & $4.87^{+0.16}_{-0.17 }$ & bad fit\\
		1870& $36.48^{+0.73}_{-0.85}$ & $3.67^{+0.09}_{-0.11}$ & $5.01^{+0.54}_{-0.54 }$ & \\
		1890& $38.52^{+0.77}_{-1.15}$ & $3.86^{+0.22}_{-0.15}$ & $5.42^{+0.05}_{-0.06 }$ & \\
		1894& $31.48^{+0.73}_{-1.11}$ & $3.5^{+0.21}_{-0.15}$ & $4.47^{+0.05}_{-0.07  }$ & \\
		1901& $33.33^{+2.04}_{-1.31}$ & $3.21^{+0.15}_{-0.11}$ & $4.73^{+0.76}_{-0.76 }$ & bad fit\\
		1912& $38.71^{+2.38}_{-1.15}$ & $4.14^{+0.24}_{-0.22}$ & $6.12^{+0.21}_{-0.2  }$ & \\
		1920& $36.63^{+1.42}_{-0.96}$ & $3.92^{+0.26}_{-0.16}$ & $5.15^{+0.91}_{-0.91 }$ & \\
		1941& $27.22^{+0.62}_{-0.77}$ & $3.99^{+0.04}_{-0.2}$ & $4.87^{+0.08}_{-0.08  }$ & bad fit\\
		1951& $31.6^{+0.65}_{-4.31}$ & $3.31^{+0.11}_{-0.12}$ & $5.02^{+0.05}_{-0.19  }$ & \\
		1952& $21.22^{+2.31}_{-0.46}$ & $3.81^{+0.14}_{-0.06}$ & $4.08^{+0.45}_{-0.44 }$ & \\
		1956& $51.55^{+0.85}_{-16.3}$ & $4.0^{+0.06}_{-1.08}$ & $5.59^{+0.83}_{-0.84  }$ & \\
		1968& $21.84^{+1.0}_{-0.62}$ & $3.8^{+0.06}_{-0.11}$ & $4.48^{+0.14}_{-0.13   }$ & \\
		1969& $35.48^{+1.96}_{-10.49}$ & $3.61^{+0.44}_{-0.2}$ & $4.88^{+0.67}_{-0.69 }$ & \\
		1973& $25.1^{+20.8}_{   }$ & $2.6^{+1.13}_{   }$ & $3.86^{+0.72}_{-0.55       }$ & grid boundary, bad fit\\
		1974& $37.33^{+0.69}_{-0.96}$ & $4.21^{+0.08}_{-0.19}$ & $5.45^{+0.04}_{-0.05 }$ & \\
		1977& $21.99^{+0.88}_{-0.73}$ & $3.81^{+0.05}_{-0.08}$ & $3.49^{+0.29}_{-0.29 }$ & \\
		1979& $37.67^{+1.23}_{-0.88}$ & $3.51^{+0.1}_{-0.08}$ & $5.43^{+0.06}_{-0.05  }$ & \\
		1988& $28.99^{+1.11}_{-1.11}$ & $4.07^{+0.05}_{-0.15}$ & $4.43^{+0.44}_{-0.44 }$ & bad fit\\
		1998& $36.06^{+1.77}_{-1.58}$ & $3.81^{+0.47}_{-0.26}$ & $4.86^{+0.16}_{-0.16 }$ & \\
		2003& $39.83^{+0.54}_{-0.58}$ & $3.71^{+0.07}_{-0.06}$ & $6.29^{+0.63}_{-0.63 }$ & \\
		2005& $45.82^{+0.73}_{-15.64}$ & $3.69^{+0.05}_{-1.03}$ & $5.03^{+0.05}_{-0.37}$ & \\
		2016& $36.6^{+0.92}_{-0.96}$ & $3.82^{+0.19}_{-0.12}$ & $5.53^{+0.2}_{-0.2    }$ & \\
		2017& $26.99^{+0.85}_{-0.62}$ & $4.01^{+0.04}_{-0.04}$ & $3.67^{+0.22}_{-0.22 }$ & \\
		2018& $21.95^{+11.19}_{-0.77}$ & $3.84^{+0.42}_{-0.09}$ & $4.26^{+0.82}_{-0.8 }$ & \\
		2024& $21.61^{+0.69}_{-0.54}$ & $3.78^{+0.05}_{-0.24}$ & $4.65^{+0.05}_{-0.04 }$ & \\
		2033& $30.6^{+1.0}_{-0.92}$ & $4.1^{+0.05}_{-0.05}$ & $5.34^{+0.44}_{-0.44    }$ & \\
		2038& $43.44^{+2.61}_{-10.26}$ & $3.51^{+0.08}_{-0.67}$ & $4.69^{+0.09}_{-0.29}$ & bad fit\\
		2044& $38.67^{+0.58}_{-0.96}$ & $4.09^{+0.23}_{-0.18}$ & $5.28^{+0.1}_{-0.1   }$ & \\
		2053& $42.44^{+2.69}_{-1.38}$ & $3.6^{+0.13}_{-0.07}$ & $5.86^{+0.11}_{-0.08  }$ & \\
		2057& $42.25^{+0.69}_{-0.85}$ & $3.49^{+0.06}_{-0.05}$ & $5.3^{+0.04}_{-0.04  }$ & bad fit\\
		2077& $44.94^{+4.23}_{-4.19}$ & $3.91^{+0.12}_{-0.08}$ & $5.81^{+0.14}_{-0.14 }$ & \\
		2086& $31.52^{+0.42}_{-1.04}$ & $3.49^{+0.07}_{-0.19}$ & $4.2^{+0.59}_{-0.59  }$ & \\
		2090& $34.6^{+1.0}_{-0.65}$ & $4.19^{+0.06}_{-0.22}$ & --                        & no photometry\\
		2097& $26.6^{+1.19}_{-3.11}$ & $3.99^{+0.06}_{-0.35}$ & $4.33^{+0.07}_{-0.17  }$ & \\
		2098& $21.8^{+0.81}_{-0.65}$ & $3.82^{+0.05}_{-0.09}$ & $4.18^{+0.65}_{-0.65  }$ & \\
		2102& $44.75^{+0.69}_{-0.77}$ & $3.8^{+0.09}_{-0.09}$ & $6.48^{+0.21}_{-0.21  }$ & \\
		2112& $46.01^{+4.38}_{-11.8}$ & $3.7^{+0.14}_{-0.77}$ & $5.7^{+0.14}_{-0.31   }$ & \\
		2114& $32.52^{+0.62}_{-0.54}$ & $3.2^{+0.06}_{-0.07}$ & $4.3^{+0.3}_{-0.3     }$ & \\
		2128& $39.98^{+1.19}_{-0.77}$ & $3.71^{+0.18}_{-0.1}$ & $5.51^{+0.53}_{-0.53  }$ & \\
		2130& $34.52^{+0.54}_{-0.54}$ & $4.09^{+0.13}_{-0.13}$ & $5.32^{+0.16}_{-0.16 }$ & \\
		2140& $18.18^{+0.27}_{-0.38}$ & $3.3^{+0.23}_{-0.14}$ & $3.26^{+0.3}_{-0.3    }$ & \\
		2147& $26.64^{+1.61}_{-3.96}$ & $3.98^{+0.08}_{-0.41}$ & $3.85^{+0.5}_{-0.51  }$ & \\
		2159& $21.37^{+0.96}_{-0.42}$ & $3.7^{+0.18}_{-0.19}$ & $3.88^{+0.6}_{-0.6    }$ & \\
		2165& $35.6^{+1.42}_{-1.04}$ & $3.5^{+0.19}_{-0.12}$ & $5.85^{+0.08}_{-0.07   }$ & \\
		2174& $22.3^{+0.62}_{-0.54}$ & $3.81^{+0.04}_{-0.06}$ & $3.91^{+0.29}_{-0.29  }$ & bad fit\\
		2176& $18.49^{+0.5}_{-0.35}$ & $3.79^{+0.04}_{-0.12}$ & $3.97^{+0.54}_{-0.54  }$ & \\
		2190& $28.22^{+0.58}_{-0.58}$ & $2.79^{+0.06}_{-0.06}$ & $5.49^{+0.12}_{-0.12 }$ & \\
		2193& $36.63^{+0.81}_{-0.69}$ & $3.79^{+0.16}_{-0.15}$ & $5.79^{+0.77}_{-0.77 }$ & \\
		2231& $38.48^{+1.58}_{-1.88}$ & $3.84^{+0.19}_{-0.13}$ & $5.41^{+0.11}_{-0.12 }$ & \\
		2233& $32.45^{+3.27}_{-1.38}$ & $4.16^{+0.07}_{-0.23}$ & $4.88^{+0.15}_{-0.07 }$ & \\
		2245& $37.75^{+1.35}_{-1.11}$ & $4.23^{+0.13}_{-0.16}$ & $5.16^{+0.06}_{-0.05 }$ & \\
		2256& $31.37^{+1.81}_{-1.77}$ & $3.48^{+0.31}_{-0.27}$ & $4.87^{+0.1}_{-0.1   }$ & \\
		2270& $29.06^{+1.77}_{-2.0}$ & $4.08^{+0.06}_{-0.06}$ & $4.75^{+0.09}_{-0.11  }$ & \\
		2301& $28.99^{+1.42}_{-3.0}$ & $4.1^{+0.04}_{-0.07}$ & $5.04^{+0.08}_{-0.15   }$ & bad fit\\
		2302& $17.8^{+0.58}_{   }$ & $3.69^{+0.03}_{-0.04}$ & $3.72^{+0.84}_{-0.84    }$ & grid boundary, bad fit\\
		2385& $54.55^{+1.65}_{-4.42}$ & $4.5^{    }_{-0.14}$ & $5.7^{+0.05}_{-0.12    }$ & grid boundary\\
		2388& $33.52^{+2.27}_{-5.11}$ & $3.2^{+0.15}_{-0.54}$ & $4.36^{+0.74}_{-0.74  }$ & \\
		2389& $31.02^{+3.84}_{-0.88}$ & $4.1^{+0.11}_{-0.18}$ & $4.93^{+0.18}_{-0.05  }$ & \\
		2395& $49.78^{+1.73}_{-3.08}$ & $4.19^{+0.22}_{-0.15}$ & $6.23^{+0.06}_{-0.1  }$ & \\
		2401& $36.67^{+1.58}_{-1.77}$ & $4.28^{+0.06}_{-0.13}$ & $4.54^{+0.53}_{-0.53 }$ & \\
		2417& $33.64^{+1.08}_{-0.88}$ & $3.24^{+0.13}_{-0.1}$ & $5.49^{+0.05}_{-0.04  }$ & \\
		2438& $18.49^{+0.5}_{-0.31}$ & $3.69^{+0.05}_{-0.07}$ & $3.67^{+0.41}_{-0.41  }$ & \\
		2447& $53.09^{+1.27}_{-9.88}$ & $4.5^{    }_{-0.45}$ & $6.04^{+0.04}_{-0.24   }$ & grid boundary\\
		2451& $44.75^{+3.84}_{-0.96}$ & $4.0^{+0.19}_{-0.15}$ & $5.46^{+0.13}_{-0.04  }$ & \\
		2453& $25.78^{+6.56}_{-0.38}$ & $2.6^{+0.31}_{0.0}$ & $3.68^{+0.39}_{-0.31    }$ & bad fit\\
		2454& $33.37^{+1.11}_{-0.73}$ & $4.19^{+0.05}_{-0.1}$ & $4.9^{+0.24}_{-0.24   }$ & bad fit\\
		2498& $50.13^{+1.19}_{-1.11}$ & $4.5^{    }_{-0.2}$ & $5.64^{+0.05}_{-0.05    }$ & grid boundary\\
		2506& $21.14^{+0.65}_{-2.88}$ & $2.9^{+0.21}_{-0.27}$ & $3.36^{+0.21}_{-0.26  }$ & \\
		2511& $26.95^{+0.88}_{-0.58}$ & $4.01^{+0.04}_{-0.04}$ & $4.18^{+0.5}_{-0.5   }$ & \\
		2521& $27.18^{+0.62}_{-0.81}$ & $3.72^{+0.24}_{-0.26}$ & $4.72^{+0.04}_{-0.05 }$ & \\
		2543& $24.8^{+1.69}_{-2.04}$ & $3.84^{+0.13}_{-0.3}$ & $4.14^{+0.17}_{-0.18   }$ & \\
		2545& $38.98^{+7.03}_{-8.99}$ & $3.17^{+0.35}_{-0.51}$ & $4.99^{+0.24}_{-0.28 }$ & bad fit\\
		2565& $35.6^{+1.23}_{-1.04}$ & $3.66^{+0.16}_{-0.13}$ & $5.09^{+0.06}_{-0.05  }$ & \\
		2570& $37.6^{+2.04}_{-1.65}$ & $4.29^{+0.08}_{-0.46}$ & $5.3^{+0.09}_{-0.07   }$ & \\
		2572& $32.25^{+3.31}_{-7.46}$ & $4.19^{+0.1}_{-0.28}$ & $4.45^{+0.29}_{-0.35  }$ & \\
		2607& $28.91^{+0.92}_{-4.77}$ & $3.99^{+0.1}_{-0.54}$ & $4.37^{+0.2}_{-0.27   }$ & \\
		2610& $28.64^{+0.81}_{-0.85}$ & $3.99^{+0.07}_{-0.36}$ & $4.53^{+0.05}_{-0.05 }$ & \\
		2653& $32.6^{+0.73}_{-0.81}$ & $3.5^{+0.13}_{-0.11}$ & $5.73^{+0.04}_{-0.04   }$ & \\
		2665& $34.67^{+1.65}_{-1.96}$ & $3.4^{+0.22}_{-0.22}$ & $4.82^{+0.08}_{-0.09  }$ & \\
		2681& $24.41^{+1.65}_{-0.65}$ & $3.32^{+0.27}_{-0.2}$ & $4.08^{+0.11}_{-0.05  }$ & \\
		2703& $30.41^{+1.11}_{-1.04}$ & $4.08^{+0.06}_{-0.28}$ & $5.07^{+0.17}_{-0.17 }$ & \\
		2718& $33.52^{+0.5}_{-0.5}$ & $3.96^{+0.14}_{-0.12}$ & $4.85^{+0.05}_{-0.05   }$ & \\
		2735& $28.37^{+1.69}_{-3.15}$ & $4.1^{+0.04}_{-0.05}$ & $4.14^{+0.42}_{-0.43  }$ & \\
		2748& $38.52^{+0.73}_{-2.54}$ & $3.61^{+0.11}_{-0.11}$ & $5.55^{+0.05}_{-0.11 }$ & \\
		2760& $39.71^{+0.88}_{-1.42}$ & $3.61^{+0.1}_{-0.11}$ & $5.64^{+0.04}_{-0.06  }$ & \\
		2763& $40.02^{+3.27}_{-0.85}$ & $3.7^{+0.11}_{-0.1}$ & $5.17^{+0.12}_{-0.04   }$ & \\
		2771& $21.49^{+0.73}_{-0.42}$ & $3.71^{+0.12}_{-0.29}$ & $4.15^{+0.11}_{-0.1  }$ & \\
		2776& $32.56^{+1.54}_{-5.3}$ & $3.2^{+0.16}_{-0.56}$ & $5.04^{+0.18}_{-0.26   }$ & \\
		2780& $37.56^{+0.69}_{-10.53}$ & $3.39^{+0.07}_{-0.83}$ & $5.41^{+0.04}_{-0.33}$ & \\
		2782& $36.52^{+0.73}_{-0.54}$ & $3.75^{+0.11}_{-0.09}$ & $4.82^{+0.04}_{-0.03 }$ & \\
		2789& $37.98^{+1.04}_{-1.19}$ & $4.31^{+0.05}_{-0.11}$ & $5.31^{+0.05}_{-0.05 }$ & \\
		2804& $24.26^{+0.46}_{-0.96}$ & $3.07^{+0.16}_{-0.15}$ & $4.49^{+0.1}_{-0.11  }$ & \\
		2813& $22.34^{+0.65}_{-0.62}$ & $3.8^{+0.06}_{-0.11}$ & $3.94^{+0.05}_{-0.05  }$ & \\
		2815& $27.52^{+1.5}_{-1.15}$ & $4.01^{+0.07}_{-0.17}$ & $4.37^{+0.41}_{-0.4   }$ & \\
		2819& $40.82^{+0.81}_{-1.38}$ & $3.62^{+0.09}_{-0.08}$ & $5.84^{+0.05}_{-0.07 }$ & \\
		2846& $33.29^{+1.19}_{-1.35}$ & $4.18^{+0.06}_{-0.44}$ & $4.36^{+0.95}_{-0.95 }$ & \\
		2848& $21.45^{+0.77}_{-0.46}$ & $3.59^{+0.24}_{-0.26}$ & $3.81^{+0.09}_{-0.08 }$ & \\
		2861& $28.06^{+1.73}_{-3.0}$ & $4.04^{+0.07}_{-0.14}$ & $4.13^{+0.33}_{-0.34  }$ & \\
		2876& $35.44^{+0.69}_{-1.23}$ & $3.85^{+0.2}_{-0.23}$ & $5.1^{+0.05}_{-0.07   }$ & \\
		2884& $38.02^{+0.96}_{-1.04}$ & $3.6^{+0.07}_{-0.07}$ & $5.14^{+0.06}_{-0.06  }$ & \\
		2893& $47.47^{+2.11}_{-2.92}$ & $4.1^{+0.26}_{-0.16}$ & $5.68^{+0.1}_{-0.12   }$ & \\
		2897& $37.44^{+1.31}_{-1.35}$ & $4.3^{+0.05}_{-0.31}$ & $5.14^{+0.06}_{-0.06  }$ & \\
		2901& $53.09^{+1.58}_{-6.53}$ & $4.5^{    }_{-0.5}$ & $5.58^{+0.05}_{-0.17    }$ & grid boundary\\
		2911& $38.67^{+1.42}_{-1.31}$ & $3.51^{+0.09}_{-0.07}$ & $5.47^{+0.61}_{-0.61 }$ & \\
		2913& $36.48^{+0.92}_{-0.85}$ & $3.96^{+0.15}_{-0.14}$ & $5.39^{+0.07}_{-0.06 }$ & \\
		2928& $22.03^{+10.76}_{-0.85}$ & $3.84^{+0.4}_{-0.11}$ & $4.15^{+0.51}_{-0.31 }$ & \\
		2929& $35.67^{+0.96}_{-0.62}$ & $3.61^{+0.16}_{-0.09}$ & $4.04^{+0.6}_{-0.6   }$ & \\
		2939& $26.64^{+1.5}_{-4.34}$ & $3.89^{+0.17}_{-0.3}$ & $4.29^{+0.67}_{-0.67   }$ & \\
		2944& $33.25^{+2.34}_{-1.77}$ & $4.19^{+0.05}_{-0.1}$ & $4.68^{+1.29}_{-1.29  }$ & \\
		2945& $36.6^{+0.73}_{-0.62}$ & $3.77^{+0.1}_{-0.11}$ & $4.92^{+0.04}_{-0.04   }$ & \\
		2946& $42.09^{+4.5}_{-3.04}$ & $3.89^{+0.23}_{-0.15}$ & $5.58^{+0.15}_{-0.11  }$ & \\
		2966& $23.07^{+4.11}_{-1.69}$ & $3.91^{+0.11}_{-0.47}$ & $3.21^{+1.18}_{-1.18 }$ & \\
		2977& $36.67^{+0.92}_{-0.65}$ & $3.8^{+0.15}_{-0.1}$ & $5.27^{+1.28}_{-1.28   }$ & \\
		2981& $38.02^{+1.04}_{-1.31}$ & $4.14^{+0.12}_{-0.18}$ & $5.47^{+0.05}_{-0.06 }$ & \\
		2985& $18.61^{+0.69}_{-0.81}$ & $3.42^{+0.22}_{-0.21}$ & $3.98^{+0.73}_{-0.73 }$ & \\
		2987& $48.44^{+3.61}_{-18.49}$ & $3.81^{+0.13}_{-1.15}$ & $5.6^{+0.62}_{-0.66 }$ & \\
		3007& $29.87^{+2.34}_{-3.08}$ & $3.3^{+0.48}_{-0.42}$ & $4.8^{+0.12}_{-0.15   }$ & bad fit\\
		3010& $44.67^{+2.88}_{-3.0}$ & $3.59^{+0.1}_{-0.15}$ & $4.89^{+0.1}_{-0.11    }$ & bad fit\\
		3018& $30.45^{+2.96}_{-4.04}$ & $4.16^{+0.07}_{-0.07}$ & $4.29^{+0.81}_{-0.81 }$ & \\
		3027& $37.71^{+3.88}_{-6.38}$ & $3.2^{+0.24}_{-0.38}$ & $4.59^{+0.15}_{-0.23  }$ & bad fit\\
		3030& $43.4^{+1.46}_{-1.85}$ & $4.4^{+0.05}_{-0.24}$ & $5.41^{+0.07}_{-0.08   }$ & \\
		3032& $25.06^{+1.85}_{-2.0}$ & $4.02^{+0.04}_{-0.05}$ & $3.71^{+0.28}_{-0.28  }$ & \\
		3034& $35.79^{+1.04}_{-0.81}$ & $3.6^{+0.1}_{-0.08}$ & $5.25^{+0.08}_{-0.07   }$ & \\
		3043& $35.56^{+1.0}_{-0.69}$ & $3.53^{+0.15}_{-0.09}$ & $5.94^{+0.05}_{-0.04  }$ & \\
		3062& $32.52^{+0.54}_{-0.5}$ & $3.2^{+0.06}_{-0.06}$ & $5.68^{+0.5}_{-0.5     }$ & \\
		3081& $37.67^{+0.58}_{-0.88}$ & $4.12^{+0.11}_{-0.22}$ & $5.36^{+0.03}_{-0.04 }$ & \\
		3106& $33.33^{+0.73}_{-0.85}$ & $4.06^{+0.16}_{-0.19}$ & $4.61^{+0.04}_{-0.05 }$ & \\
		3129& $22.18^{+4.23}_{-1.19}$ & $3.8^{+0.24}_{-0.1}$ & $3.44^{+0.29}_{-0.2    }$ & \\
		3149& $22.34^{+0.65}_{-0.92}$ & $3.8^{+0.07}_{-0.1}$ & $4.06^{+0.51}_{-0.52   }$ & \\
		3151& $22.18^{+0.69}_{-0.96}$ & $3.83^{+0.08}_{-0.17}$ & $3.78^{+0.97}_{-0.97 }$ & bad fit\\
		\hline
\end{longtable}
\end{center}
\end{spacing}
\twocolumn                                

\bsp	
\label{lastpage}
\end{document}